\documentclass[sigconf]{acmart}

\setcopyright{none}
\usepackage{ulem} %to strike the words
\usepackage{enumitem}
\usepackage{threeparttable} 
\usepackage{multirow}
\usepackage{amsmath}
\usepackage{mathrsfs}
\usepackage{graphicx}
\usepackage[T1]{fontenc}
\usepackage{aecompl}
\usepackage{subfigure}

\def \W{{\bf W}}
\def \b{{\bf b}}
\def \x{{\bf x}}

\renewcommand\footnotetextcopyrightpermission[1]{} % removes footnote with conference information in first column

\AtBeginDocument{%
  }

\begin{document}

\title{CTRL: Connect Collaborative  and Language Model for CTR Prediction }


% \author{%
% Xiangyang Li$^{*}$, 
% Bo Chen$^{*}$, 
% Lu Hou, 
% Ruiming Tang
% }
% \thanks{*Co-first authors with equal contributions.}

% \affiliation{Huawei Noah's Ark Lab\\
% \{lixiangyang34, chenbo116, houlu3, tangruiming\}@huawei.com}

\author{Xiangyang Li}
\authornote{Co-first authors with equal contributions. }
\email{lixiangyang34@huawei.com}
\affiliation{%
\country{China}
  \institution{Huawei Noah's Ark Lab}
}

\author{Bo Chen}
 \authornotemark[1]
\email{chenbo116@huawei.com}
\affiliation{%
\country{China}
  \institution{Huawei Noah's Ark Lab}
}

\author{Lu Hou}
\email{houlu3@huawei.com}
\affiliation{%
\country{China}
  \institution{Huawei Noah's Ark Lab}
}

\author{Ruiming Tang}
\email{tangruiming@huawei.com}
\affiliation{%
\country{China}
  \institution{Huawei Noah's Ark Lab}
}

\renewcommand{\shortauthors}{Xiangyang Li and Bo Chen et al.}
\begin{abstract}
Traditional click-through rate (CTR) prediction models convert the tabular data into one-hot vectors and leverage the collaborative relations among features for inferring the user's preference over items. 
This modeling paradigm discards essential semantic information. 
Though some works like P5 and CTR-BERT have explored the potential of using Pre-trained Language Models (PLMs) to extract semantic signals for CTR prediction, they are computationally expensive and suffer from low efficiency. 
Besides, the beneficial collaborative relations are not considered, hindering the recommendation performance.
To solve these problems, in this paper, we propose a novel framework \textbf{CTRL}, which is industrial-friendly and model-agnostic with superior inference efficiency. 
Specifically, the original tabular data is first converted into textual data. Both tabular data and converted textual data are regarded as two different modalities and are separately fed into the collaborative CTR model and pre-trained language model. 
A cross-modal knowledge alignment procedure is performed to fine-grained align and integrate the collaborative and semantic signals, and the lightweight collaborative model can be deployed online for efficient serving after fine-tuned with supervised signals. 
Experimental results on three public datasets show that CTRL outperforms the state-of-the-art (SOTA) CTR models significantly. Moreover, we further verify its effectiveness on a large-scale industrial recommender system. 
% Code is available
% at: https://anonymous.4open.science/r/CTRL\_123/.
% Code is available
% at: https://anonymous.4open.science/r/CTRL\_123/.
% Our code and data will be open-sourced in this repository\footnote{https://gitee.com/mindspore/models/tree/master/research/recommend} \cb{code?}.

\end{abstract}

\maketitle

\section{Introduction}

Click-through rate (CTR) prediction is an important task for recommender systems and online advertising~\cite{graepel2010web,mcmahan2013ad}, where users' willingness to click on items is predicted based on historical behavior data. 
The estimated CTR is leveraged to determine whether an item can be displayed to the user.
Consequently, accurate CTR prediction service is critical to improving user experience, product sales, and advertising platform revenue~\cite{zhang2014optimal}.

For the CTR prediction task, historical data is organized in the form of tabular data.
During the evolution of recommendation models, from the early Matrix Factorization (MF)~\cite{mf}, to shallow machine learning era models like Logistic Regression (LR)~\cite{LR} and Factorization Machine (FM)~\cite{FM}, and continuing to the deep neural models such as DeepFM~\cite{guo2017deepfm} and DIN~\cite{din}, \textbf{collaborative signals} have always been the core of recommendation modeling, which leverages the feature co-occurrences and label signals for inferring user preferences.
After encoding the tabular features into one-hot features~\cite{he2014practical}, the co-occurrence relations (i.e., interactions) of the features are captured by various human-designed operations (e.g., inner product~\cite{pnn,guo2017deepfm}, outer product~\cite{cfm,fgcnn}, non-linear layer~\cite{zhang2016deep,widedeep}, etc.). By modeling these collaborative signals explicitly or implicitly, the relevance between users and items can be inferred.
\vspace{-2.0em}
\begin{figure}[htbp]
\centering
\includegraphics[scale=0.35]{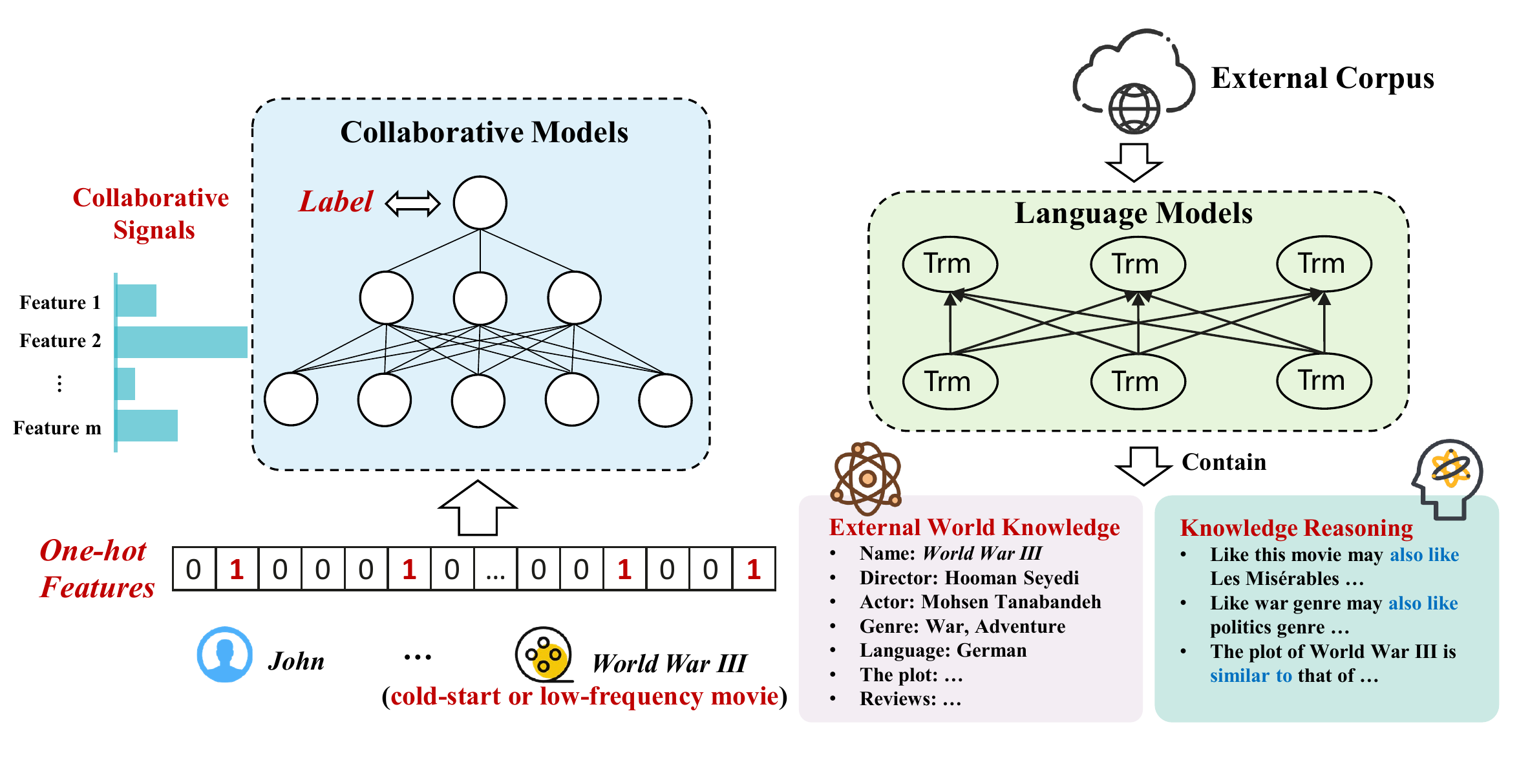}
\caption{\small{The external world  knowledge and reasoning capabilities of pre-trained language models facilitate recommendations.}}
\label{motivation}
\vspace{-1.5em}

\end{figure}

However, the collaborative-based modeling paradigm discards the semantic information among the original features due to the one-hot feature encoding process. 
Therefore, for cold-start scenarios or low-frequency long-tailed features, the recommendation performance is unsatisfactory, limited by the inadequate collaborative relations~\cite{lu2020meta}.
For example, in Figure~\ref{motivation}, when inferring the click probability of user \texttt{John} over a cold start movie \texttt{World War \uppercase\expandafter{\romannumeral3}}, the inadequate collaborative signals in historical data may impede accuracy recommendation. 
Recently, some works are proposed to address this drawback by involving Pre-trained Language Models (PLMs) to model \textbf{semantic signals}, such as P5~\cite{p5}, M6-Rec~\cite{m6-rec}, CTR-BERT~\cite{ctr-BERT}, TALLRec~\cite{tallrec}, PALR~\cite{palr}. 
These works feed the original textual features directly into the language models for recommendation, rather than using one-hot encoded features.
On the one hand, the linguistic and semantic knowledge in PLMs helps to extract the semantic information within the original textual features~\cite{ptab}. On the other hand, the \textit{external world knowledge} such as the director, actors, even story plot and reviews for the movie \texttt{World War \uppercase\expandafter{\romannumeral3}}, as well as \textit{knowledge reasoning capability} in Large Language Models (LLMs) provide general knowledge beyond training data and scenarios~\cite{zhang2021language}, thus enlightening a new technological path for recommender systems. 
% WeChat proposes InstructRec
% ~\cite{InstructRec}, which introduces LLM into recommendation systems by instruct tuning.

% , but provides \textit{external world knowledge}, such as the director, actors, even story plot and reviews for the movie \texttt{World War \uppercase\expandafter{\romannumeral3}}, thus exploring a new technological path for the recommender systems. 

% For example, ``\texttt{judge}'' and ``\texttt{lawyer}'' are semantically close professions but are regarded as completely independent features after one-hot encoding. linguistic

% Recently, some works are proposed to address this drawback by involving Pre-trained Language Models (PLM) to model \textbf{semantic signals}, such as P5~\cite{p5}, M6-Rec~\cite{m6-rec}, CTR-BERT~\cite{ctr-BERT}. These works feed the original textual features rather than one-hot encoded features into the language models for extracting semantic information and perform recommendation directly.

Although remarkable progress has been achieved, the existing semantic-based solutions suffer from several shortcomings: 1) 
% \sout{They explore a different approach to make prediction with semantics by abandoning the traditional collaborative modeling, which is suboptimal. Because the feature co-occurrence patterns are an indispensable indicator for personalized recommendation. Discarding them may degrade the performance.}
Making predictions based on semantics merely without traditional collaborative modeling can be suboptimal~\cite{p5} because the feature co-occurrence patterns and user-item interactions are indispensable indicators for personalized recommendation~\cite{guo2017deepfm}, which are not yet well equipped for PLMs~\cite{zhang2021language, liu2023chatgpt}.
2) 
% \sout{Deploying language models for online serving brings tremendous pressure on online inference latency due to sophisticated architectures.}
Online inferences of language models are computationally expensive due to their complex structures.
To adhere to low-latency constraints, massive computational resources, and engineering optimizations are involved, hindering large-scale industrial applications~\cite{m6-rec,p5}.

Therefore, incorporating PLMs into recommendation systems to capture semantic signals confronts two major challenges: 
\begin{itemize}[leftmargin=*]
\item How to combine the collaborative signals with semantic signals to boost the performance of recommendation?
\item How to ensure efficient online inference without involving extensive engineering optimizations?
\end{itemize}

To solve these two challenges above, inspired by the recent works in contrastive learning, we propose a novel framework to \textbf{C}onnec\textbf{t} Collabo\textbf{r}ative and \textbf{L}anguage Model (\textbf{CTRL}) for CTR prediction, which consists of two stages:  \textbf{Cross-modal Knowledge Alignment} stage, and \textbf{Supervised Fine-tuning} stage.
Specifically, the raw tabular data is first converted into textual data by human-designed prompts, which can be understood by language models. Then, the original tabular data and generative textual data are regarded as different modalities and fed into the collaborative CTR model and pre-trained language model, respectively. We execute a cross-modal knowledge alignment procedure, meticulously aligning and integrating collaborative signals with semantic signals. Finally, the collaborative CTR model is fine-tuned on the downstream task with supervised signals. During the online inference, only the lightweight fine-tuned CTR model is pushed for serving without the language model, thus ensuring efficient inference.

Our main contributions are summarized as follows:
\vspace{-0.05cm}
\begin{itemize}[leftmargin=*]
\item We first propose a novel training framework CTRL, which is capable of aligning signals from collaborative and language models,  introducing semantic knowledge into the collaborative models.

\item Through extensive experiments, we demonstrate that the incorporation of semantic knowledge significantly enhances the performance of collaborative models on CTR task.
% \item CTRL treats the tabular data and textual data as two modalities and leverages the contrastive learning for fine-grained knowledge alignment and integration, thus providing adequate modeling capability for collaborative and semantic signals. 
\item CTRL is industrial-friendly, model-agnostic, and can adapt with any collaborative models and PLMs, including LLMs. Moreover, the high inference efficiency is also retained, facilitating its application in industrial scenarios.
\item In experiments conducted on three publicly available datasets from real-world industrial scenarios,  CTRL achieved SOTA performance. Moreover, we further verify its effectiveness on large-scale industry recommender systems.
\end{itemize}

\section{Related Work}
% In this section, we will discuss the connections and differences between CTRL and existing collaborative and semantic models.

\subsection{Collaborative Models for Recommendation}
\label{collaborative}

During the evolution of recommendation models, from the early matrix factorization (MF)~\cite{mf}, to shallow machine learning era models like Logistic Regression (LR)~\cite{LR} and Factorization Machine (FM)~\cite{FM}, to the deep neural models~\cite{guo2017deepfm,din}, collaborative signals have always been the core of recommendation modeling. These collaborative-based models convert the tabular features into one-hot features and leverage various interaction functions to extract feature co-occurrence relations (a.k.a. feature interactions).

Different human-designed interaction functions are proposed to improve the modeling ability of collaborative signals. Wide\&Deep~\cite{widedeep} uses the non-linear layers to extract implicit high-order interactions.
DeepFM~\cite{guo2017deepfm} leverages the inner product to capture pairwise interactions with stacked and parallel structures. 
% CFM~\cite{cfm} uses the convolution operation to identify the local feature interaction patterns.
DCN~\cite{dcn} and EDCN~\cite{edcn} deploy cross layers to model bit-wise feature interactions.
% Moreover, some AutoML-based CTR models are proposed to search suitable feature interactions and interaction functions, such as AIM~\cite{aim}, and AutoFeature~\cite{autofeature}.

Though collaborative-based models have achieved significant progress, they cannot capture the semantic information of the original features, thereby hindering the prediction effect in some scenarios such as cold-start or low-frequency long-tailed features. 

\vspace{-1.5em}
% Although knowledge graph-based recommendation models~\cite{kgat} can extract some semantics, the captured information is limited because the dependent knowledge graph is generally for a certain vertical field.
% Plenty of features are not included in the knowledge graph, thus limiting the performance.
% Therefore, introducing a general semantic modeling method is vital and language models provide a sterling approach.

%基于协同的ctr prediction,知识图谱

\subsection{Semantic Models for Recommendation}
%基于语义的推荐模型ctr-BERT，ptab，p5，m6-rec,
%

Transformer-based language models, such as BERT~\cite{BERT}, GPT-3~\cite{gpt-3}, and T5~\cite{t5}, have emerged as foundational architectures in the realm of Natural Language Processing (NLP).
% Typically, these models undergo pre-training on voluminous web text data and subsequent fine-tuning on downstream tasks~\cite{transformer}. 
Their dominance across various NLP subdomains, such as text classification~\cite{text-1,text-2}, sentiment analysis~\cite{sa-1,sa-2}, intelligent dialogue~\cite{p5,instruct-gpt}, and style transfer~\cite{transfer-1,transfer-2}, is primarily attributed to their robust capabilities for knowledge reasoning and transfer.  Nevertheless, since recommender systems mainly employ tabular data, which is heterogeneous with text data, it is difficult to apply the language model straightforwardly to the recommendation task.   

In recent times, innovative research trends have surfaced, exploring the viability of language models in recommendation tasks. P5~\cite{p5}, serves as a generative model tailored for recommendations, underpinning all downstream recommendation tasks into a text generation task and utilizing the T5~\cite{t5} model for training and prediction. P-Tab~\cite{ptab} introduces a recommendation methodology based on discriminative language models, translating tabular data into prompts, pre-training these prompts with a Masked Language Model objective, and finally fine-tuning on downstream tasks. Concurrently, Amazon's CTR-BERT~\cite{ctr-BERT}, a two-tower structure comprising two BERT models, encodes user and item text information respectively. More recently, a considerable upsurge in scholarly works has been observed, leveraging Large Language Models (LLMs) for recommendation systems~\cite{gpt-search,gpt-rank,tallrec,llm-fair,InstructRec}. For instance, a study by Baidu~\cite{gpt-search} investigates the possibility of using LLM for re-ranking within a search context. Similarly, RecLLM~\cite{llm-fair} addresses the issue of fairness in the application of LLMs within recommendation systems. 
However, although the above semantic-based recommendation models have exposed the possibility of application in recommender systems, they have two fatal drawbacks: 
1) Discarding the superior experience accumulation in collaborative modeling presented in Section~\ref{collaborative} and making prediction with semantics only may be suboptimal~\cite{p5} and hinder the performance for cold-start scenarios or low-frequency long-tailed features.
2) Due to the huge number of parameters of the language models, it is quite arduous for language models to meet the low latency requirements of recommender systems, making online deployment much more challenging. Instead, our proposed CTRL overcomes these two shortcomings by combining both collaborative and semantic signals via two-stage training paradigm.

% by combining both collaborative and semantic signals via pre-training multimodal comparisons, which allows it not only to have excellent performance but also to meet low latency requirements.
% recommend systems.

%
\section{preliminary}
%现有推荐系统无法利用好语义信息，抽象出一个现有推荐模型的范式（one hot embedding），没有利用好文本信号

In this section, we present the collaborative-based deep CTR model and reveal the deficiencies in modeling semantic information.
The CTR prediction is a supervised binary classification task, whose dataset consists of several instances $(\mathbf{x}, y)$. Label $y \in \{0, 1\}$ indicates user’s actual click action. Feature $\mathbf{x}$ is multi-fields that contains important information about the relations between users and items, including user profiles (e.g., gender, occupation), item features (e.g., category, price) as well as contextual information (e.g., time, location)~\cite{autodis}. Based on the instances, the traditional deep CTR models leverage the collaborative signals to estimate the probability $P(y=1|\mathbf{x})$ for each instance.

The existing collaborative-based CTR models first encode the tabular features into one-hot features and then model the feature co-occurrence relations by various human-designed operations. Specifically, the multi-field tabular features are transformed into the high-dimensional sparse features via field-wise one-hot encoding~\cite{he2014practical}. For example, the feature (Gender=\texttt{Female},  Occupation=\texttt{Doctor}, Genre=\texttt{Sci-Fi}, \dots, City=\texttt{Hong Kong}) of an instance can be represented as a one-hot vector:
\begin{equation}
\label{onehot}
\small
    \mathbf{x} = \underbrace{[0,1]}_{\text{Gender}} \underbrace{[0,0,1,\dots,0]}_{\text{Occupation}}
    \underbrace{[0,1,0,\dots,0]}_{\text{Genre}}
    \dots
    \underbrace{[0,0,1,\dots,0]}_{\text{City}}.
\end{equation}

Generally, deep CTR models follow an ``Embedding \& Feature interaction'' paradigm~\cite{edcn,autodis}. The high-dimensional sparse one-hot vector is mapped into a low-dimensional dense space via an embedding layer with an embedding look-up operation. Specifically, for the $i$-th feature, the corresponding feature embedding $\mathbf{e}_i$ can be obtained via $\mathbf{e}_i = \mathbf{E}_i \mathbf{x}_i$, where $\mathbf{E}_i$ is the embedding matrix. Following, feature interaction layers are proposed to capture the explicit or implicit feature co-occurrence relations. Massive effort has been made in designing specific interaction functions, such as product~\cite{pnn,guo2017deepfm}, cross layer~\cite{dcn,edcn,xdeepfm}, non-linear layer~\cite{zhang2016deep,widedeep}, and attention layer~\cite{din}.
Finally, the predictive CTR score $\hat{y}$ is obtained via an output layer and optimized with the ground-truth label $y$ through the widely-used Binary Cross Entropy (BCE).

As we can observe, collaborative-based CTR models leverage the one-hot encoding to convert the original tabular data into one-hot vectors as E.q.(\ref{onehot}), discarding the semantic information among the feature fields and values\footnote{We use ``feature field'' to represent a class of features following~\cite{autodis} and ``feature value'' to represent a certain value in a specific field. For example, \texttt{occupation} is a ``feature field'' and \texttt{doctor} is one of the ``feature value''.}.
By doing this, the feature semantics is lost and the only signals that can be used for prediction are the feature co-occurrence relations, which is suboptimal when the relations are weak in some scenarios such as cold-start or low-frequency long-tailed features.
Therefore, introducing the language model to capture the essential semantic information is conducive to compensating for the information gaps and improving performance.

\section{METHOD}

% \begin{figure*}[htbp]
% \centering
% \includegraphics[scale=0.45]{figure/model_8.pdf}
% \caption{\small{An intuitive illustration of the three-stage paradigm of the CTRL. CTRL consists of  three stages: Prompt Transfer, Multimodal Contrastive Learning, and Supervised Fine-tuning.}}
% \label{model}
% \end{figure*}

\begin{figure*}[htbp]
	\centering
	\setlength{\belowcaptionskip}{-0.3cm}
	\setlength{\abovecaptionskip}{0cm}
	\begin{tabular}{c}
		\includegraphics[scale=0.3]{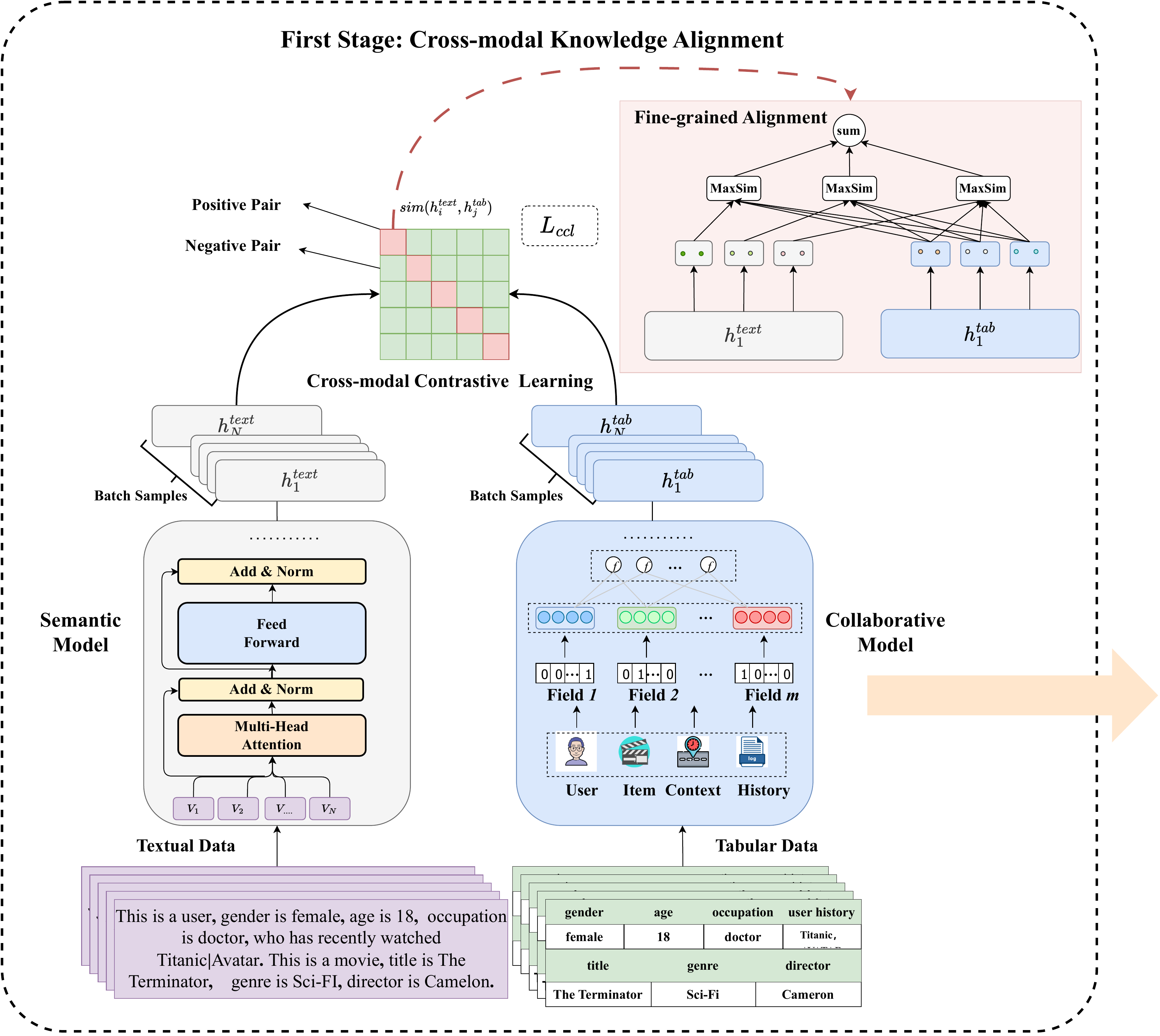} 
		\hspace{-0.7em}
		\includegraphics[scale=0.3]{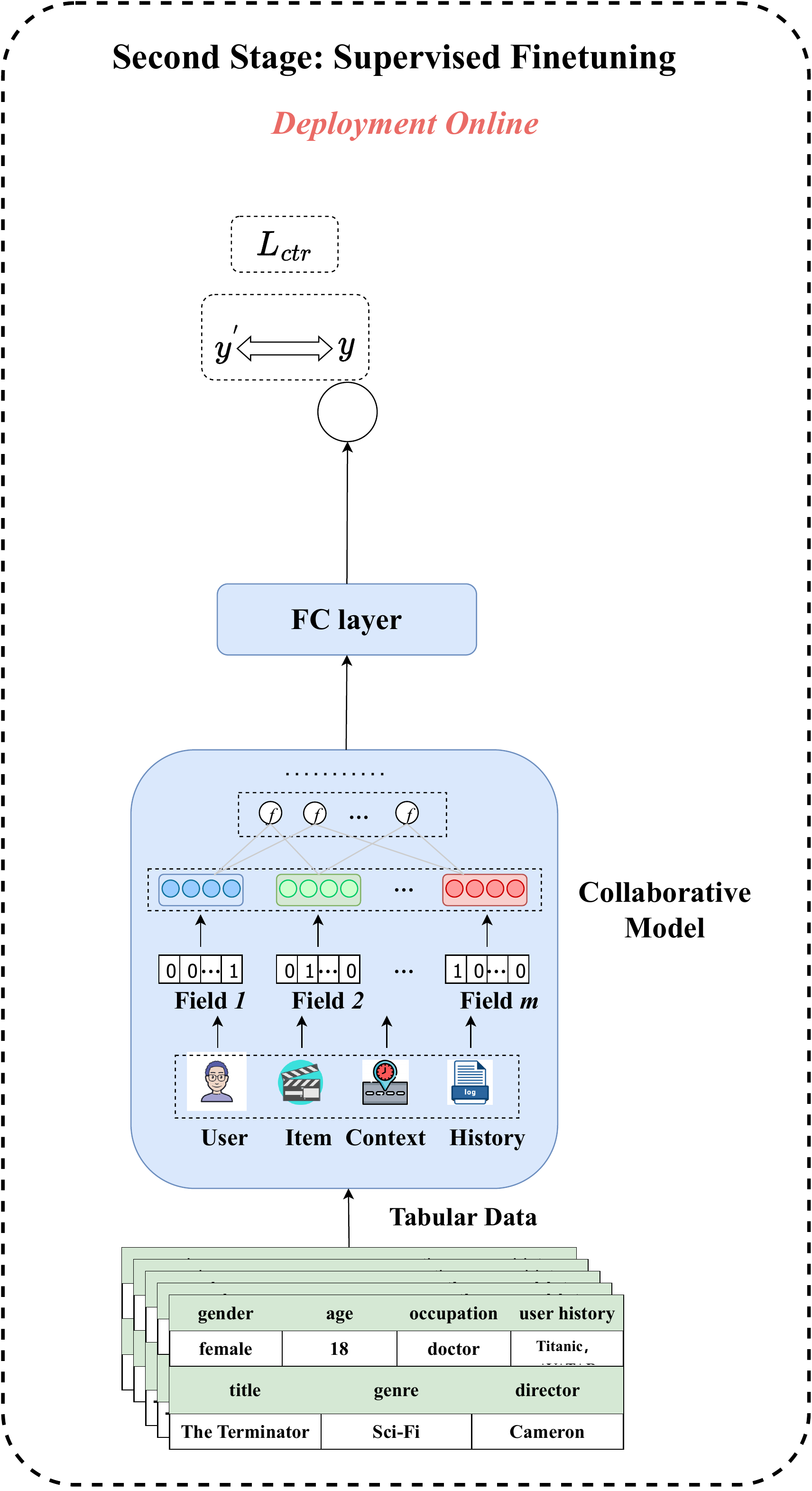}
	\end{tabular}
	\caption{\small{An intuitive illustration of the CTRL, which is a two-stage framework, in the first stage, cross-modal contrastive learning is used to fine-grained align knowledge of the two modalities. In the second stage, the lightweight collaborative model is fine-tuned on downstream tasks. The red square represents a positive pair in the batch, while the green square represents a negative pair.
% 	The red squares in the figure represent the calculation of positive case similarity, whereas the green squares pertain to that of negative examples.
	}}
	\label{model}

\end{figure*}

%模型的总览
As depicted in Figure~\ref{model}, the proposed CTRL is a two-stage training paradigm. 
% The first one is the \textbf{Prompt Construction} stage, which converts the tabular data into prompts (i.e., text data) that can be identified by the language model. 
The first stage is \textbf{Cross-modal Knowledge Alignment}, which feeds paired tabular data and textual data from two modalities into the collaborative model and the language model respectively, and then aligns them with the contrastive learning objective. 
The second stage is the \textbf{Supervised Fine-tuning} stage, where the collaborative model is fine-tuned on the downstream task with supervised signals.

\vspace{-0.5em}
\begin{figure}[htbp]
\centering
\includegraphics[scale=0.4]{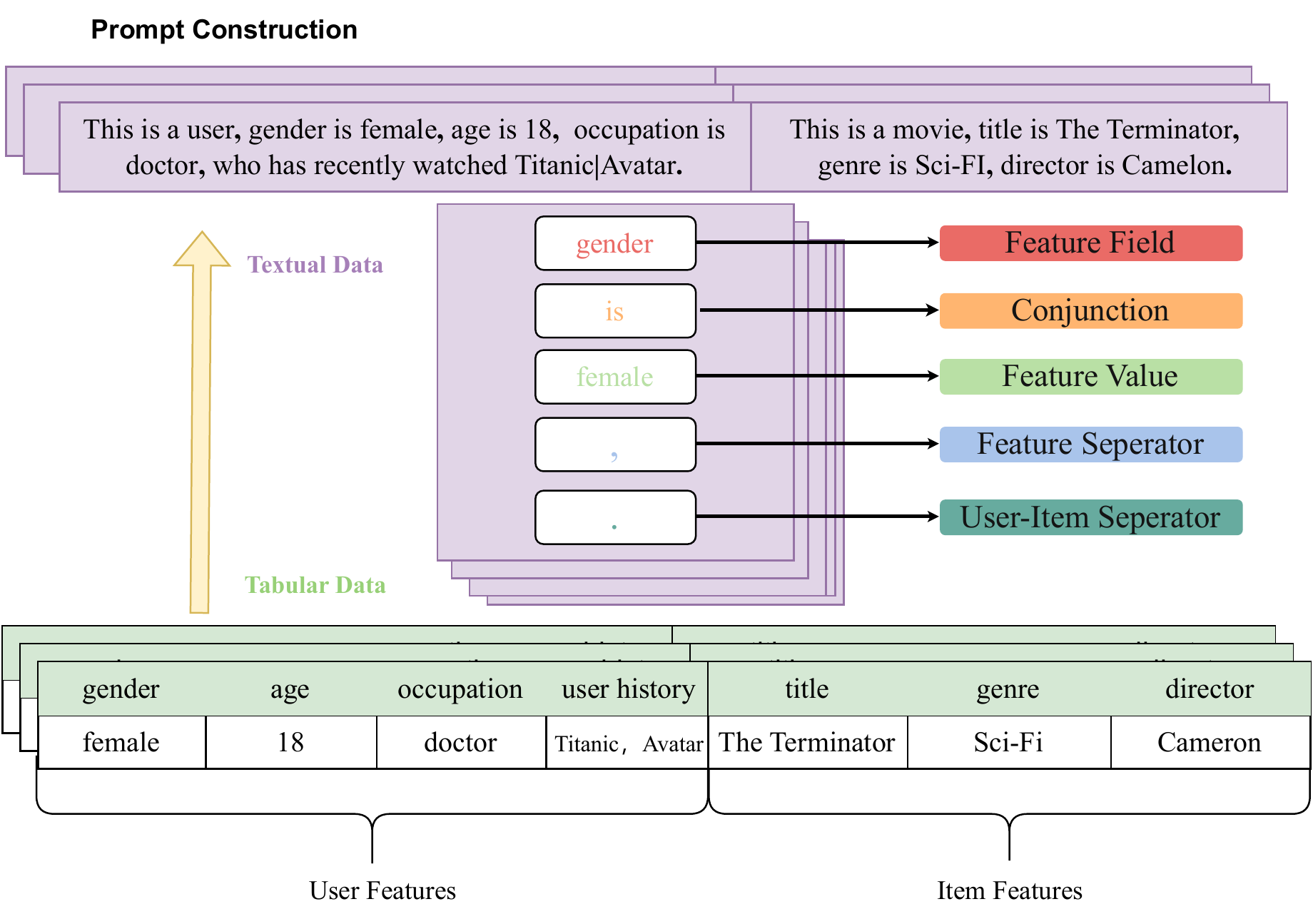}
\caption{\small{The overall process of prompt construction.}}
\label{prompt_cons}
\vspace{-1.9em}
\end{figure}

\subsection{Prompt Construction}
%模板如何构建(prompt)
% As shown in Figure~\ref{model} stage 1, to  map paired texts and tabular data to the same multimodal space, 
% we create personalized  prompt templates. 

% As shown in Figure~\ref{prompt_cons}, during the first stage, CTRL first create personalized  prompt templates to covert the tabular data into text data for each instance.

% \begin{figure*}[htbp]
% 	\centering
% 	\setlength{\belowcaptionskip}{-0.3cm}
% 	\setlength{\abovecaptionskip}{0cm}
% 	\subfigure[\small{First Stage}]{
% 		\includegraphics[scale=0.3]{figure/ctrl_stage1.pdf}}
%  	\hspace{-3.5em}
% 	\subfigure[\small{Second Stage}]{
% 		\includegraphics[scale=0.3]{figure/second_stage.pdf}}
% 	\caption{\small{An intuitive illustration of the the working flow of the CTRL framework. CTRL is a two-stage framework, where in the first stage, we use contrastive learning pre-training to align the spaces of the two modalities. In the second stage, we fine-tune the lightweight collaborative model on downstream tasks.}}
% 	\label{model}
% \end{figure*}

% \begin{figure*}[htbp]
% \centering
% \includegraphics[scale=0.33]{figure/ctrl_4.pdf}
% \caption{\small{An intuitive illustration of the the working flow of the CTRL framework. CTRL is a two-stage framework, where in the first stage, we use contrastive learning pre-training to align the spaces of the two modalities. In the second stage, we fine-tune the pre-trained collaborative model on downstream tasks.}}
% \label{model}
% \end{figure*}

Before introducing the two-stage training paradigm, we first present the prompt construction process.
As illustrated in Figure~\ref{prompt_cons}, to obtain textual prompt data,  we design prompt templates to transform the tabular data into textual data for each training instance. As mentioned in previous work~\cite{p5,m6-rec}, a proper prompt should contain sufficient semantic information about the user and the item. 
For example, user's profiles such as age, identity, interests, and behaviors can be summarized in a single sentence. Besides, item's description sentence can be organized with the features such as color, quality, and shape. 
For this purpose, we design the following template to construct the prompts:

\begin{quote}
This is a user,  gender is female,  age is 18, occupation is doctor,  who has recently watched Titanic|Avatar. This is a movie, title is The Terminator, genre is Sci-FI,   director is Camelon.
\end{quote}

% \begin{quote}
% [BOU]This is a user, user id is 1, gender is Female,  occupation is doctor, age is 30,  who has recently watched Titanic|Avatar.[EOU] [BOI] This is a Movie, title is The Terminator, genre is Sci-FI,  country of production is the United States,  film length is 120 minutes, director is Camelon.[EOI]
% \end{quote}
In this prompt,  
the first sentence ``\texttt{This is a user,  gender is female,  age is 18,  occupation is doctor, who has recently watched Titanic|Avatar.}''
% the text between [BOU] and [EOU] 
describes the user-side features,
including his/her profiles such as age, gender, occupation, and historical behaviors, etc. 
% The text between [BOI] and [EOI]  
The following sentence ``\texttt{This is a movie, title is The Terminator, genre is Sci-FI,  director is Camelon.}'' describes
the item-side features such as title, category, director, etc. In the practical implementation, we use the period \textbf{``.''} to separate the user-side and item-side descriptions, the comma \textbf{``,''} to separate each feature, and vertical bar \textbf{``|''} to separate each user's historical behavior\footnote{Note that this step is performed in the data process pipeline, and generating millions of textual prompts only takes a few seconds with parallel computing. For datasets with hundreds of features, a subset of significant features is selected to generate prompts.}. 
We also explore the effect of different prompts, of which results are presented in Section~\ref{sec:ablation_prompt}.

% \begin{figure*}[htbp]
% \centering
% \includegraphics[scale=0.62]{figure/prompt.pdf}
% \caption{\small{Building text input from raw tabular data according to our designed personalized prompt templates.}}
% \label{IntTower}

% \end{figure*}
\subsection{Cross-modal Knowledge Alignment}
As mentioned before, existing collaborative-based recommendation models~\cite{dcn,autoint} leverage the feature co-occurrence relations to infer users' preferences over items, facilitating the evolution of recommendations. Besides, the pre-trained language models~\cite{BERT} specializes in capturing the semantic signals of recommendation scenarios with the linguistic and external world knowledge~\cite{p5}. In order to combine the modeling capabilities of both collaborative-based models and pre-trained language models, as well as ensure efficient online inference, CTRL proposes an implicit information integration method via contrastive learning~\citep{simclr,gao2021simcse}, where cross-modal knowledge (i.e., tabular and textual information) between collaborative and semantic space is aligned. 
\vspace{-0.8em}
%With this, the information of the two modals can be fully fused, 

% In the previous step, we created personalized prompts that allow us to create pairs of text and tabular data efficiently. 
% In this stage, we need to map texts and tabular data
% into the same multimodal embedding space, which facilitates  collaborative model to learn the semantic representation of language model. 
% Contrastive learning aims at pushing paired samples close and apart otherwise, and is proved effective in both unimodal~\citep{simclr,gao2021simcse} and cross-modal \citep{clip,filip} representation learning. 
% Inspired by these works, CTRL also applies the contrastive learning objective, to align 
% the textual and tabular information into a unified semantic space. 

\subsubsection{Cross-modal  Contrastive Learning}
 The cross-modal contrastive procedure is presented in Figure~\ref{model}. First, the collaborative model and semantic model (a.k.a., pre-trained language model) are utilized to encode the tabular and textual data for obtaining the corresponding representations, respectively. Specifically, let $\mathcal{M}_{col}$ denotes collaborative model, and $\mathcal{M}_{sem}$ denotes semantic model, for an instance $\mathbf{x}$, $\mathbf{x}^{tab}$ denotes the tabular form, and $\mathbf{x}^{text}$ denotes the textual form of the same instance that is obtained after the prompt construction process.
The instance representations under collaborative and semantic space can be presented as $\mathcal{M}_{col}(\x^{tab})$ and $\mathcal{M}_{sem}(\x^{text})$, respectively. To convert the unequal-length representations into the same dimension, a linear projection layer is designed, and the transformed instance representations can be obtained as follows:
 \begin{equation}
\label{e1}
\begin{aligned}
\mathbf{h}^{tab} =  \mathcal{M}_{col}(\x^{tab}) \W^{tab}+\b^{tab} ,
\end{aligned}
\end{equation}
 \begin{equation}
\label{e1}
\begin{aligned}
\mathbf{h}^{text} = \mathcal{M}_{sem}(\x^{text}) \W^{text}+\b^{text} ,
\end{aligned}
\end{equation}
where $\mathbf{h}^{tab}$ and $\mathbf{h}^{text}$ are the transformed collaborative and semantic representations for the same instance $\x$, $\W^{tab}, \W^{text}$ and  $\b^{tab}, \b^{text}$ are the transform matrices and biases of the linear projection layers.

Then, the contrastive learning is used to align the instance representations under different latent spaces, which is proved effective in both unimodal~\citep{simclr,gao2021simcse} and cross-modal \citep{clip} representation learning. The assumption behind this is that, under a distance metric, the correlated representations should be constrained to be close, and vice versa should be far away.  We employ InfoNCE~\cite{infonce} to align two representations under collaborative and semantic space for each instance. As shown in Figure~\ref{model}, two different modalities (textual, tabular) of the same sample form a positive pair. Conversely, data from two different modalities (textual and tabular) belonging to diverse samples form a negative pair. Negative pairs are obtained through in-batch sampling.
Denote $\mathbf{h}^{text}_k, \mathbf{h}^{tab}_k$ are the representations of two modals for the $k$-th instance, the textual-to-tabular contrastive loss can be formulated as:
\begin{equation}
\begin{aligned}
\mathcal{L}^{textual2tabular}= -\frac{1}{N}\sum_{k=1}^N log\frac{exp(sim(\mathbf{h}^{text}_k, \mathbf{h}^{tab}_k)/ \tau )}{\sum_{j=1}^N exp(sim(\mathbf{h}^{text}_k, \mathbf{h}^{tab}_j)/ \tau )},
\end{aligned}
\end{equation}
where $\tau$ is a temperature coefficient and $N$ is the number of instances in a batch. Besides, function $sim(\cdot, \cdot)$ measures the similarity between two vectors. Typically, cosine similarity is employed for this purpose.
% \begin{equation}
% \begin{aligned}
% \label{sim}
% sim(\mathbf{h}_i, \mathbf{h}_j) = \frac{\mathbf{h}_i^\top \mathbf{h}_j}{\|\mathbf{h}_i\| \cdot \|\mathbf{h}_j\|}.
% \end{aligned}
% \end{equation}
In order to avoid spatial bias towards collaborative modal, motivated by the Jensen–Shannon (J-S) divergence~\cite{js}, we also design a tabular-to-textual contrastive loss for uniformly aligning into a multimodal space, which is shown as:
\begin{equation}
\begin{aligned}
\mathcal{L}^{tabular2textual}= -\frac{1}{N}\sum_{k=1}^N log\frac{exp(sim(\mathbf{h}^{tab}_k, \mathbf{h}^{text}_k)/ \tau )}{\sum_{j=1}^N exp(sim(\mathbf{h}^{tab}_k, \mathbf{h}^{text}_j)/ \tau )}.
\end{aligned}
\end{equation}

Finally, the cross-modal contrastive learning loss $\mathcal{L}_{ccl}$ is defined as the average of $\mathcal{L}^{textual2tabular}$ and $\mathcal{L}^{tabular2textual}$, and all the parameters including collaborative model $\mathcal{M}_{col}$ and semantic model $\mathcal{M}_{sem}$ are trained.
\begin{equation}
\label{ssl}
\mathcal{L}_{ccl}= \frac{1}{2}(\mathcal{L}^{textual2tabular} + \mathcal{L}^{tabular2textual}).
% \label{eq:contrastive_loss}
\end{equation}

\subsubsection{Fine-grained Alignment}
As mentioned above, CTRL leverages the cross-modal contrastive learning to perform knowledge alignment, where the quality of alignment is measured by the cosine similarity function. 
However, this approach models the global similarities merely and ignores fine-grained information alignment between the two modalities $\mathbf{h}^{tab}$ and $\mathbf{h}^{text}$. To address this issue, CTRL adopts a fine-grained cross-modal alignment method.

Specifically, both collaborative and semantic representations $\mathbf{h}^{tab}$ and $\mathbf{h}^{text}$ are first transformed into  $M$ sub-spaces to extract informative knowledge from different aspects. Taking the collaborative representation $\mathbf{h}^{tab}$ as example, the $m$-th sub-representation $\mathbf{h}_{m}^{tab}$ is denoted as:
 \begin{equation}
\label{sub_space_u}
\begin{aligned}
\mathbf{h}_{m}^{tab}=\mathbf{W}^{tab}_{m}\mathbf{h}^{tab}+\mathbf{b}^{tab}_{m}, \ \ \ \ \ \ \ m=1,2,\dots,M,
\end{aligned}
\end{equation}
where $\mathbf{W}^{tab}_{m}$ and $\mathbf{b}^{tab}_{m}$ are the transform matrix and bias vector for the $m$-th sub-space, respectively. Similarly, the $m$-th sub-representation for semantic representation is denoted as $\mathbf{h}_{m}^{text}$.

Then, the fine-grained alignment is performed by calculating the similarity score, which is conducted as a sum of maximum similarity over all sub-representations, shown as:
  \begin{equation}
\label{sub_space_text}
\begin{aligned}
 sim(\mathbf{h}_i, \mathbf{h}_j) = \sum_{m_i = 1}^{M} \max \limits _{m_j \in 1,2,\dots, M}  \{  (\mathbf{h}_{i,m_i})^T \mathbf{h}_{j,m_j}\},
\end{aligned}
\end{equation}
where $\mathbf{h}_{i,m}$ is the $m$-th sub-representation for representation $\mathbf{h}_{i}$. 
By modeling fine-grained similarity over the cross-modal spaces, CTRL allows for more detailed alignment within instance representations to better integrate knowledge. In this stage, both the language model and collaborative model parameters are updated to better align the representations.

\vspace{-0.2cm}

\subsection{Supervised Fine-tuning}
% \begin{figure}[htbp]
% \centering
% \includegraphics[scale=0.3]{figure/sft.pdf}
% \caption{\small{An intuitive illustration of supervised fine-tuning process.}}
% \label{sft}
% \end{figure}

After the cross-modal knowledge alignment stage, the collaborative knowledge and semantic knowledge are aligned and aggregated in a hybrid representation space, where the relations between features are mutually strengthened.
In this stage, CTRL further fine-tunes the collaborative models on different downstream tasks (CTR prediction task in this paper) with supervised signals.

% After the large-scale pre-training in the previous stage, the representations of the collaborative and semantic models have been jointly mapped into the same multimodal representation space.
% In this stage, we fine-tune the collaborative model on downstream tasks. After pre-training, the collaborative model allows fine-tuning on multiple recommendation tasks. In this work, we mainly investigate performance on the CTR prediction task.

At the top of the collaborative model, we add an extra linear layer with random initialization, acting as the output layer for final prediction $\hat{y}$. The widely-used Binary Cross Entropy (BCE) loss is deployed to measure the classification accuracy between the prediction score $\hat{y}$ and the ground-truth label $y$, which is defined as follows:
 \begin{equation}
\label{e1}
\begin{aligned}
\mathcal{L}_{ctr} = -\frac{1}{N}\sum_{k=1}^{N}(y_{k}log(\hat{y}_{k}) + (1-y_{k})log(1-\hat{y}_{k}) ),
\end{aligned}
\end{equation}
where $y_{k}$ and $\hat{y}_{k}$ are the ground-truth label and the model prediction score of the $k$-th instance.
After the supervised fine-tuning stage, only the lightweight collaborative model will be deployed online for serving, thus ensuring efficient online inference.

\section{Experiments}
% In this section, we describe the experiments in detail, including the experimental settings, comparison with the SOTA baseline models and the corresponding analysis. Through experiments such as performance comparisons and efficiency studies, we aim to answer the following research questions about our proposed CTRL framework:
% \begin{itemize}[leftmargin=*]
% 	\item RQ1: How does CTRL perform compared to the SOTA models?
% 	\item RQ2: Can CTRL meet the requirement of low inference latency, which is vital for industrial recommender systems?
% 	\item RQ3: How well CTRL aligns the collaborative and semantic spaces, which reflects the effect of knowledge integration.
% % 	Does CTRL really align the semantic signal space and the collaborative signal space? How does the performance vary with different prompts?
% % 	\item RQ4: Is the language model sufficiently competent for the CTR task and is it possible to surpass the collaborative model?
% 	\item RQ4: Does CTRL have sufficient compatibility? To what extent does it affect the performance when applying with different semantics and collaborative models, including LLMs?
% 	\item RQ5: Can CTRL be applied in large-scale industrial scenarios?
% % 	\item RQ5: If CTRL perform well on public datasets, will CTRL work well on industrial level systems with sufficient performance improvements?
	
% \end{itemize}

\subsection{Experimental Setting}

\subsubsection{Datasets and Evaluation Metrics}
In the experiment, we deploy three large-scale public datasets, which are MovieLens, Amazon (Fashion), and Taobao, whose statistics are summarized in Table~\ref{Table:stat.}.
Following previous work~\cite{din,autoint,huang2019fibinet}, we use two popular metrics to evaluate the performance, i.e., \textbf{AUC} and \textbf{Logloss}. 
% A lower bound of 0 for Logloss indicates that the two distributions are perfectly matched, and a smaller value indicates a better performance.  
As acknowledged by many studies~\cite{din,autoint,fibinet}, an improvement of \textbf{0.001} in AUC ($\uparrow$) or Logloss ($\downarrow$) can be regarded as significant because it will bring a large increase in the online revenue.
\textbf{RelaImpr} metric~\cite{din} measures the relative improvement with respect to base model, which is defined as follows: 
 \begin{equation}
\label{e1}
\begin{aligned}
RelaImpr = (\frac{AUC(measure\ model)-0.5}{AUC(base\ model)-0.5} -1)\times 100\%.
\end{aligned}
\end{equation}
Besides, the two-tailed unpaired $t$-test is performed to detect a significant difference between CTRL and the best baseline.
The detailed description of datasets and metrics can be referred to Appendix~\ref{appendix_setting}.

\begin{table}[!t]

\footnotesize

\caption{\small{Basic statistics of datasets.}}
 \resizebox{.47\textwidth}{!}{
\begin{tabular}{@{}cccccc@{}}
\toprule
Dataset         & Users       & Items  & User Field & Item Field & Samples    \\ \midrule
MovieLens-1M  & 6,040      & 3,952  & 5        & 3     & 1,000,000    \\
Amazon(Fashion)      & 749,232      & 196,637  & 2      & 4     & 883,636  \\
Alibaba         & 1,061,768 & 785,597 & 9       & 6    & 26,557,961 \\ \bottomrule

\end{tabular}
}
\label{Table:stat.}
\vspace{-2.0em}
\end{table}

% \subsubsection{Evaluation Metrics}

% It is worth noting that  a slightly higher AUC or lower Logloss at \textbf{0.001-level} is regarded as significant for CTR prediction task, which has also been pointed out by previous works~\cite{din,autoint}.
% The two-tailed unpaired $t$-test~\cite{bhattacharya2002median} is performed to detect significant differences between CTRL and the best baseline.
\subsubsection{Competing Models}
%与传统的Ctr模型进行比较，deepfm，dcn，autoint
%与预训练语言模型进行比较，P5，ctr-BERT
We compare CTRL with the following models, which are classified into two classes, i.e.,  1) Collaborative Models:\textbf{Wide\&Deep}~\cite{widedeep}, \textbf{DeepFM}~\cite{guo2017deepfm}, \textbf{DCN}~\cite{dcn}, 
\textbf{PNN}~\cite{pnn}, \textbf{AutoInt}~\cite{autoint}, 
\textbf{FiBiNet}~\cite{fibinet}, and \textbf{xDeepFM}~\cite{xdeepfm}; and 2) Semantic Models: \textbf{P5}~\cite{p5}, \textbf{CTR-BERT}~\cite{ctr-BERT}, and \textbf{P-Tab}~\cite{ptab}. 
The detailed description of these models can be referred to Appendix~\ref{appendix_models}.

\begin{table*}[!h]
% \normalsize
\setlength\tabcolsep{5pt}
\caption{\small{Performance comparison of different models. 
%The base model of RelaImpr is DSSM \cb{remove RelaImpr}, which is a popular neural network-based recommendation model. 
The boldface denotes the highest score and the underline indicates the best result of all baselines. $\star$ represents significance level $p$-value $<0.05$ of comparing CTRL with the best baselines. RelaImpr denotes the relative AUC improvement rate of CTRL against each baseline.}}

\begin{threeparttable}[!t]
\small
\begin{tabular}{@{}ccccccccccccc@{}}
\toprule
\multirow{2}{*}{Category}  & \multirow{2}{*}{Model} & \multicolumn{4}{c}{MovieLens} & \multicolumn{4}{c}{Amazon} & \multicolumn{3}{c}{Alibaba} \\ 
\cmidrule(l){3-13} 
& & AUC & Logloss & RelaImpr & & AUC & Logloss & RelaImpr & & AUC & Logloss & RelaImpr \\ 
\cmidrule(l){1-13} 
\multirow{7}{*}{Collaborative  Models}
% & DSSM & 0.7901 & 0.4826 &16.37\%  & & 0.6481 & 0.4815 &40.04\% & & 0.5696 & 0.3559 & 92.24\% \\  
& Wide\&Deep & 0.8261 & 0.4248 &3.52\% & & 0.6968 & 0.4645 &5.30\% & & 0.6272 & 0.1943 &5.19\% \\
& DeepFM & 0.8268 & 0.4219 &3.30\% & & 0.6969 & 0.4645 &5.33\% & & 0.6280 & 0.1951 &4.53\% \\
& DCN &\underline{0.8313} & \underline{0.4165} &1.90\%  & & 0.6999 & 0.4642 &3.75\% & & \underline{0.6281} & 0.1949 &4.45\% \\ 
& PNN &0.8269 & 0.4220 &3.27\%  & & 0.6979 & 0.4657 &4.80\% & & 0.6271 & 0.1956 &5.27\% \\ 
& AutoInt & 0.8290 & 0.4178 &2.61\% & & \underline{0.7012} & \underline{0.4632} &3.08\% & & 0.6279 & \underline{0.1948} &4.61\% \\
& FiBiNet & 0.8196 & 0.4188 &5.63\% & & 0.7003 & 0.4704 &3.54\% & & 0.6270 & 0.1951 &5.35\% \\
& xDeepFM & 0.8296 & 0.4178 &2.43\% & & 0.7009 & 0.4642 &3.23\% & & 0.6272 & 0.1959 &5.19\% \\
\midrule
\multirow{3}{*}{Semantic Models}   
& P5 & 0.7583 & 0.4912 &30.70\% & & 0.6923 & 0.4608 &7.85\% & & 0.6034 & 0.3592 &29.40\% \\
& CTR-BERT & 0.7650 & 0.4944 &27.40\% & & 0.6934 & 0.4629 &7.24\% & & 0.6005 & 0.3620 &33.13\% \\
& P-Tab & 0.8031 & 0.4612 &11.38\% & & 0.6942 & 0.4625 &6.80\% & & 0.6112 & 0.3584 &20.32\% \\
\midrule
\multicolumn{2}{c}{CTRL} & \textbf{0.8376}$^\star$ & \textbf{0.4025}$^\star$ &- & & \textbf{0.7074}$^\star$ & \textbf{0.4577}$^\star$ &- & & \textbf{0.6338}$^\star$ & \textbf{0.1890}$^\star$ &- \\

% \multicolumn{2}{c}{Rel Impr}. & 0.76\% & 3.36\% & & & 0.88\% & 1.18\% & & & 0.91\% & 2.97\% & \\
\bottomrule
\end{tabular}
\begin{tablenotes}    
\huge \item[*] \small \textbf{It is worth noting that an AUC increase of 0.001 can be considered a significant  improvement
in CTR prediction~\cite{inttower,din,autoint,fibinet}.
}
\end{tablenotes} 
\end{threeparttable}
\vspace{-1.4em}
\label{performance}
\end{table*}

\subsubsection{Implementation Details}
%We implement the proposed CTRL and other compared baselines with Pytorch~\cite{NEURIPS2019_9015} and 
% For CTR-BERT, we feed user prompt and item prompt into the two tower model composed of BERT. For P5, we take user-item prompt as input, and the target is "positive" if  label is 1 and "negative" if  label is 0. 

For the prompt construction process, only one type of prompt is used and the comparisons are presented in Section~\ref{sec:ablation_prompt}. In the first stage, we utilize AutoInt~\cite{autoint} as the collaborative model and RoBERTa~\cite{roBERTa} as the semantic model by default, as discriminative language models are more efficient at text representation extraction than generative models like GPT under the same parameter scale~\cite{superglue}. Additionally, we also evaluated the performance of the \textbf{LLM} model like ChatGLM, with the results summarized in Table~\ref{modelsize}.
The mean pooling results of the last hidden states are used as the semantic information representation. For the projection layer, we compress the collaborative representation and the semantic representation to 128 dimensions. Besides, the batch size of the cross-modal knowledge alignment stage is set to 6400 and the temperature coefficient is set to 0.7. The AdamW~\cite{adamw} optimizer is used and the initial learning rate is set to $1\times10^{-5}$, which is accompanied by a warm-up mechanism~\cite{resnet} to 
$5\times10^{-4}$. In the second stage, the learning rate of the downstream fine-tuning task is set to 0.001 with Adam~\cite{https://doi.org/10.48550/arxiv.1412.6980} optimizer, and batch size is set to 2048. Batch Normalization~\cite{bn} and Dropout~\cite{dropout} are also applied to avoid overfitting. The feature embedding dimension $d$ for all models is set to 32 empirically. Besides, for all collaborative models, we set the number of hidden layers $L$ as 3 and the number of hidden units as $[256,128,64]$. To ensure a fair comparison, other hyperparameters such as training epochs are adjusted individually for all models to obtain the best results. %We conduct experiments with 24 Tesla V100 GPUs.

\vspace{-1.2em}

\subsection{Performance Comparison}

We compare the overall performance with some SOTA collaborative and semantic models, whose results are summarized in Table~\ref{performance}. From this, we obtain the following observations: 
1) CTRL outperforms all the SOTA baselines including semantic and collaborative models over three datasets by a significant margin, showing superior prediction capabilities and proving the effectiveness of the paradigm of combining collaborative and semantic signals.
2) In comparison to the best collaborative model, our proposed CTRL achieves an improvement in AUC of \textbf{1.90\%}, \textbf{3.08\%}, and \textbf{4.45\%} on the three datasets respectively, which effectively demonstrates that integrating semantic knowledge into collaborative models contributes to boost performance.
We attribute the significant improvements to the external world knowledge and knowledge reasoning capability in PLMs~\cite{zhang2021language}.
3) The performance of existing semantic models is lower than that of collaborative models, indicating that collaborative signals and co-occurrence relations are crucial for recommender systems, and relying solely on semantic modeling is difficult to surpass the existing collaborative-based modeling scheme\cite{p5,ctr-BERT,ptab}. %Additionally, there is a significant performance gap between the generative-based model P5 and the discriminative-based model CTR-BERT. 
Instead, our proposed CTRL integrates the advantages of both by combining collaborative signals with semantic signals for recommendation. This approach is likely to be a key path for the future development of recommender systems.

\vspace{-0.18cm}
\begin{table}[h]
\caption{\small{Inference efficiency comparison of different models in terms of Model Inference Parameters and Inference Time over testing set with single V100 GPU.  As for CTRL, only the collaborative model is needed for online serving, so the number of model parameters is the same as the backbone AutoInt.}} %For the model with two-tower structure, we only calculate the parameters of the user tower, since only the user tower is required for online inference.}}
\resizebox{0.85\linewidth}{!}{
\begin{tabular}{@{}ccccc@{}}
\toprule
               & \multicolumn{2}{c}{Alibaba} & \multicolumn{2}{c}{Amazon} \\ \midrule
Model & Params      & Inf Time       & Params        & Inf Time      \\ \midrule
% DSSM             &    6.71$\times 10^7$         &     15s            &    3.35$\times 10^7$        &   0.51s           \\
DeepFM             &    8.82$\times 10^7$         &     18s            &    3.45$\times 10^7$        &   0.58s           \\
 DCN              &  8.84$\times 10^7$           &    19s             &     3.46$\times 10^7$       &      0.59s        \\
 AutoInt             &    8.82$\times 10^7$         &  19s              &      3.45$\times 10^7$         &     0.59s          \\ 
P5            &    2.23$\times 10^8$         &  10832s              &       1.10$\times 10^8$         &    440s            \\
CTR-BERT            &    1.10$\times 10^8$         &  4083s               &      1.10$\times 10^8$         &     144s          \\ 
\textbf{CTRL(ours)}            &    8.82$\times 10^7$         &  19s              &      3.45$\times 10^7$         &     0.59s \\ 
 \bottomrule
 
\end{tabular}}
\label{runtime}
\vspace{-2.5em}

\end{table}

\subsection{Serving Efficiency}
%比较上线的参数，发现control和推荐模型大小一致，并且远远小于P5，ctr-BERT
In industrial recommender systems, online model serving has a strict limit, e.g., 10$\sim$20 milliseconds. Therefore, high service efficiency is essential for CTR models. In this section, we compare the model parameters and inference time of different CTR models over the Alibaba and Amazon datasets, shown in Table~\ref{runtime}.

We can observe that existing collaborative-based CTR models have fewer model parameters and higher inference efficiency in comparison with semantic-based models. Moreover, the majority of parameters for the collaborative-based models are concentrated in the embedding layer while the hidden network has very few parameters, thus benefiting the online serving. 
On the contrary, the semantic-based models (e.g., P5 and CTR-BERT), have a larger number of parameters and lower inference efficiency due to the complex Transformer-based structures, hindering the industrial applications. 
Instead, for the CTRL with AutoInt as skeleton models, both model parameters and inference time are the same as the original AutoInt model, which is thanks to the decoupled training framework (semantic model is not required for online inference) and ensures the high online serving efficiency.

% In the case of CTRL, since we perform the prediction only using the collaborative model, it has both the performance of SOTA and low latency, which is important for \textbf{industrial applications}.

% Without doubt, DSSM have the fastest inference speed and the lowest parameters due to its simple structure. The semantic-based models CTR-BERT and P5, on the contrary, have the largest number of parameters and inference latency due to the large size of the pre-trained language model and the complex computational complexity of transformer architecture. Especially for P5, the number of parameters and the inference latency are several orders of magnitude higher than 
% collaborative models. In the case of CTRL, since we perform the prediction only using the collaborative model, it has both the performance of SOTA and low latency, which is important for \textbf{industrial applications}.

\vspace{-0.3cm}

\subsection{Visualization of Modal Alignment}
\label{visualization}

\begin{figure}[htbp]
\centering
\subfigure[Before Alignment]{  %小图题的名称
\includegraphics[scale=0.39]{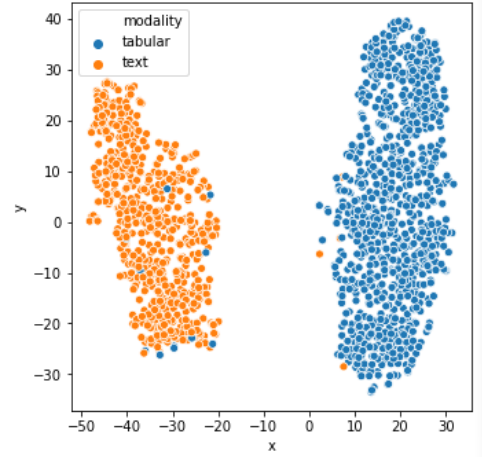}}
\hspace{0in}
\subfigure[After Alignment]{
\includegraphics[scale=0.39]{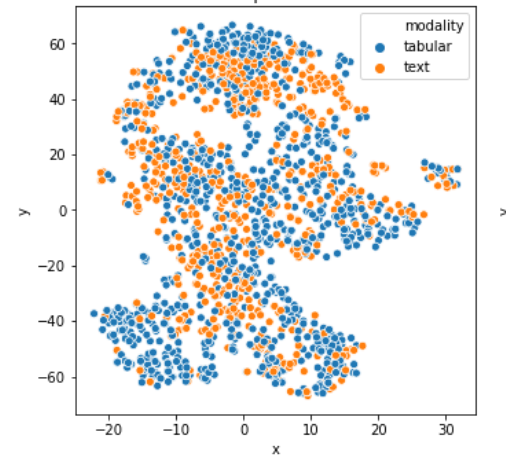}}
\caption{\small{Visualization of the tabular and textual representations before and after the cross-modal knowledge alignment.}}
\label{tsne}
\vspace{-2.0em}
\end{figure}

To study in depth the distribution of tabular representations and textual representations in the latent space before and after the cross-modal knowledge alignment, we visualize the representations in the MovieLens dataset by projecting them into a two-dimensional space using t-SNE~\cite{t-SNE}, shown in Figure~\ref{tsne}. The two colored points represent the tabular and textual representations, respectively. We can observe that, before the cross-modal knowledge alignment, the representations of the two modalities are distributed in two separate spaces and are essentially unrelated, while mapped into a unified multimodal space after the alignment.
This phenomenon substantiates that CTRL  aligns the space of two modalities (i.e., tabular and textual), thus injecting the semantic information and external general knowledge into the collaborative model.

\vspace{-1.0em}

\subsection{Compatibility Study}
%In this section, we discuss its compatibility.
\subsubsection{Compatibility for semantic models}
Specifically, for semantic models, we compare four pre-trained language models with different sizes: TinyBERT~\cite{tinyBERT} with 14.5M parameters ($\text{CTRL}_\text{TinyBERT}$), BERT-Base~\cite{BERT} with 110M parameters ($\text{CTRL}_\text{BERT}$), RoBERTa~\cite{roBERTa} with 110M parameters  ($\text{CTRL}_\text{RoBERTa}$), and BERT-Large with 336M parameters ($\text{CTRL}_\text{Large}$). Moreover, we have introduced a novel LLM model, ChatGLM~\cite{glm}, with 6B parameters ($\text{CTRL}_\text{ChatGLM}$). For $\text{CTRL}_\text{ChatGLM}$, during the training process,  we 
freeze the majority of the parameters and only retain the parameters of the last layer.
The experimental results are summarized in Table~\ref{modelsize}, from which we obtain some observations: 
1) In comparison with the backbone model AutoInt, CTRL with different pre-trained language models achieves consistent and significant improvement, where AUC increases by \textbf{3.22\%} and \textbf{3.63\%} for $\text{CTRL}_\text{ChatGLM}$ , demonstrating the effectiveness of semantics modeling and model compatibility.
2) Among the four CTRL variants ($\text{CTRL}_\text{TinyBERT}$, $\text{CTRL}_\text{BERT}$, and $\text{CTRL}_\text{BERTLarge}$, $\text{CTRL}_\text{ChatGLM}$), despite a substantial number of parameters being frozen in ChatGLM, $\text{CTRL}_\text{ChatGLM}$ achieves optimal performance. This phenomenon indicates that enlarging the size of the language model can imbue the collaborative model with a wealth of worldly knowledge. Furthermore, even when the parameter scale of the language model is elevated to the billion level, it continues to make a positive contribution to the collaborative model. 3) It can be observed that while the parameter size of ChatGLM is several times that of BERTLarge, the gains are 
mild. Therefore, when conducting modality alignment, it is only necessary to select language models of moderate scale, such as RoBERTa.
% As for the phenomenon of performance decline in CTRL-L, we assume that the imbalanced parameters between collaborative model and semantics model may weaken the effect of knowledge alignment, thus hindering the prediction accuracy.
4) Using only TinyBert can lead to a 0.005 increase in AUC, indicating that we can use lightweight pre-trained language models to accelerate model training. 4) $\text{CTRL}_\text{RoBERTa}$ has a better performance in the case of an equal number of parameters compared to $\text{CTRL}_\text{BERT}$. We hypothesize that this improvement is due to RoBERTa possessing a broader range of world knowledge and a more robust capability for semantic modeling compared to BERT. This indirectly underscores the advantages of increased knowledge in facilitating the knowledge alignment process in collaborative models.

% In this section, we discuss the compatibility of CTRL from two perspectives: (1)The impact of changing the size of the pre-trained language model on the performance. (2) The impact of changing the different collaborative models on the performance.

% We train CTRL  with four different pre-trained language models. CTRL-S, employing TinyBERT, 14.5M parameters with 4 transformers layers. CTRL-B, employing BERT-Base, 110M parameters with 12 transformer layers. CTRL-R, employing RoBERTa~\cite{roBERTa}, 110M parameters with 12 transformer layers. CTRL-L, employing  BERT-Large, 220M parameters with 24 transformer layers. 

% The experimental results are summarized in Table
% ~\ref{modelsize}, from which we obtain some observations: (1) CTRL-B shows better performance in comparison to CTRL-S, indicating that an appropriate increase in the size of the language model is beneficial to improve the level of knowledge incorporation
% (2) CTRL-R has a better performance in the case of equal number of parameters compared to CTRL-B, which we speculate is attributed to the fact that RoBERTa contains more semantic knowledge and is capable of enhancing the effects of distillation.
% (3) CTRL-L is weaker than CTRL-S, which we assume is due to the fact that excessive knowledge injection makes the collaborative model less capable of capturing the collaborative signals. Nevertheless, CTRL-L still outperforms the collaborative model, demonstrating that there is a significant effect enhancement by injecting knowledge from the language model into the collaborative model.

\begin{table}[htbp]
\vspace{-1.0em}
\caption{\small{Model compatibility study with different semantic models.}}
\resizebox{0.85\linewidth}{!}{
\begin{tabular}{@{}ccccc@{}}
\toprule
               & \multicolumn{2}{c}{MovieLens} & \multicolumn{2}{c}{Amazon} \\ \midrule
Model & AUC      & Logloss        & AUC        & Logloss       \\ \midrule
% DeepFM             &    0.8268         &     0.4219            &    0.6965       &   0.4606            \\
%  DCN              &  0.8313           &    0.4165             &     0.6999       &      0.4642        \\
 AutoInt (backbone)            &    0.8290         &  0.4178               &      0.7012         &     0.4632            \\ 
$\text{CTRL}_\text{TinyBERT}$ (14.5M)          &    0.8347       &  0.4137               &       0.7053        &     0.4612           \\
$\text{CTRL}_\text{BERT}$ (110M)           &    0.8363         &  0.4114               &      0.7062        &     0.4609            \\ 
$\text{CTRL}_\text{RoBERTa}$  (110M)          &    0.8376         &  0.4025               &      0.7074        &     0.4577            \\ 
$\text{CTRL}_\text{BERTLarge}$ (336M)           &    0.8380        &  0.4040               &      0.7076        &     0.4574            \\ 
$\text{CTRL}_\text{ChatGLM}$ (6B)           &    \textbf{0.8396}        &  \textbf{0.4010}               &      \textbf{0.7085}        &     \textbf{0.4537}            \\ 
 \bottomrule
 
\end{tabular}}
\label{modelsize}
\vspace{-1.0em}

\end{table}

\begin{table}[htbp]
\vspace{-0.5em}
\caption{\small{Model compatibility study with different collaborative models. The semantic model is set to RoBERTa. }}
\resizebox{0.75\linewidth}{!}{
\begin{tabular}{@{}ccccc@{}}
\toprule
               & \multicolumn{2}{c}{MovieLens} & \multicolumn{2}{c}{Amazon} \\ \midrule
Model & AUC      & Logloss        & AUC        & Logloss       \\ \midrule
Wide\&Deep             &    0.8261        &     0.4348            &    0.6966        &   0.4645            \\
$\text{CTRL}_\text{Wide\&Deep}$            &    0.8304        &  0.4135               &      0.7001         &     0.4624            \\
\midrule
 DeepFM              &  0.8268           &    0.4219             &     0.6965       &      0.4646        \\
 $ \text{CTRL}_\text{DeepFM}$           &    0.8305         &  0.4136               &       0.7004         &     0.4625            \\
 \midrule
 DCN             &    0.8313         &  0.4165               &      0.6999         &     0.4642            \\ 
 $\text{CTRL}_\text{DCN}$            &    0.8365         &  0.4029              &      0.7055         &     0.4615            \\ 
 \midrule
  AutoInt             &    0.8290         &  0.4178               &      0.7012         &     0.4632            \\ 
$\text{CTRL}_\text{AutoInt}$            &    0.8376         &  0.4025              &      0.7063         &     0.4582            \\ 
 \bottomrule
\end{tabular}}
\vspace{-1.2em}
\label{diff_model}
\end{table}

\subsubsection{Compatibility for collaborative models}
Besides, we apply  CTRL to different collaborative models, including Wide\&Deep, DeepFM, DCN, and AutoInt.
From Table~\ref{diff_model}, we can observe that CTRL achieves remarkable improvements with different collaborative models consistently. The average improvements over
RelaImpr metric are \textbf{1.31\%} for Wide\&Deep, \textbf{1.13\%} for DeepFM, \textbf{1.57\%} for
DCN, and \textbf{2.61\%}  for AutoInt respectively, which demonstrates the effectiveness and model compatibility.

\subsection{Ablation Study}

\subsubsection{Ablation Study Analysis.} 
\begin{figure}[htbp]
   \vspace{-2.5em}
	\centering
	\setlength{\belowcaptionskip}{-0.3cm}
	\setlength{\abovecaptionskip}{0cm}
	\subfigure[\small{Movielens}]{
		\includegraphics[scale=0.14]{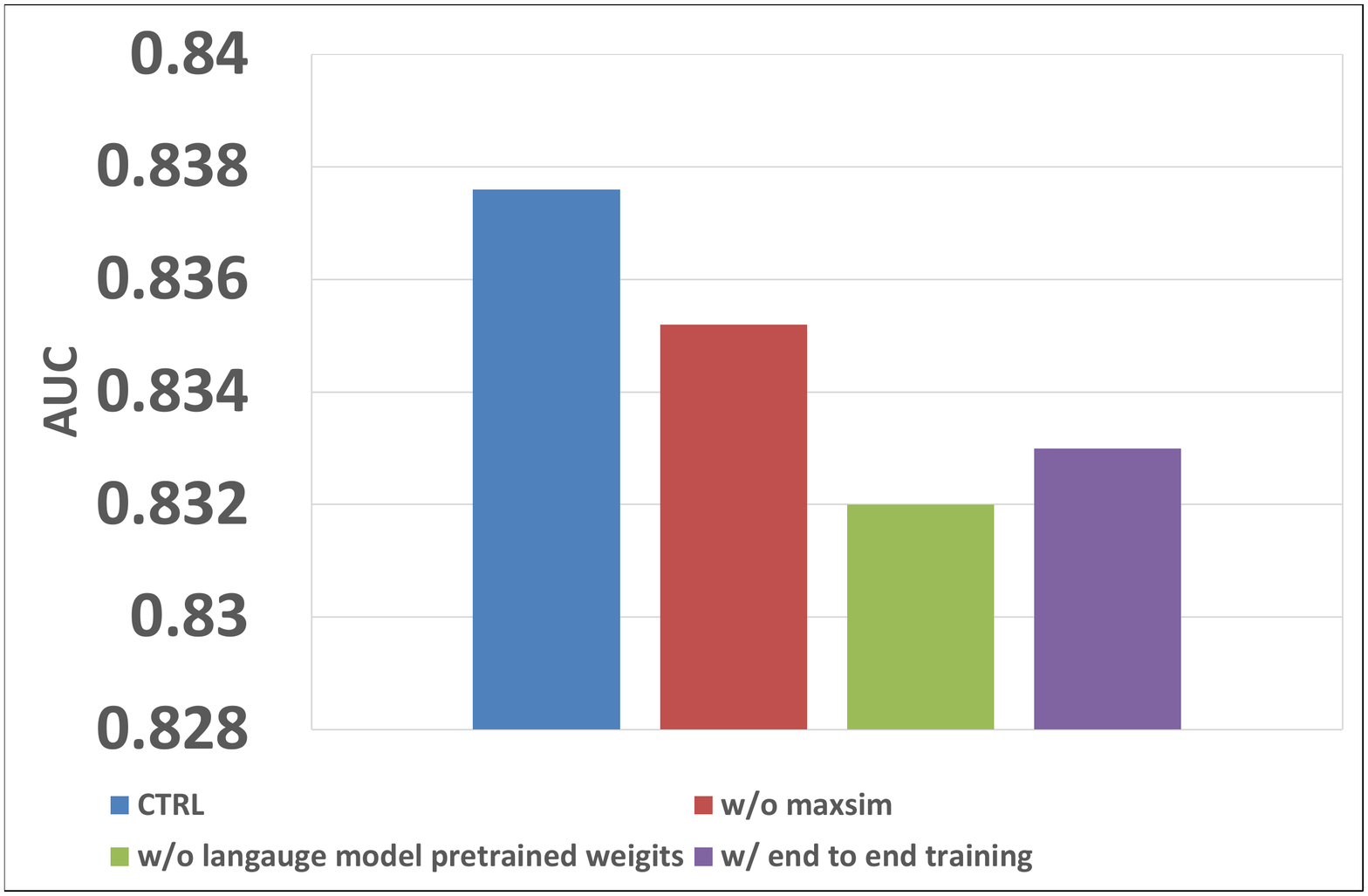}}
	\hspace{-2em}
	\subfigure[\small{Amazon}]{
		\includegraphics[scale=0.14]{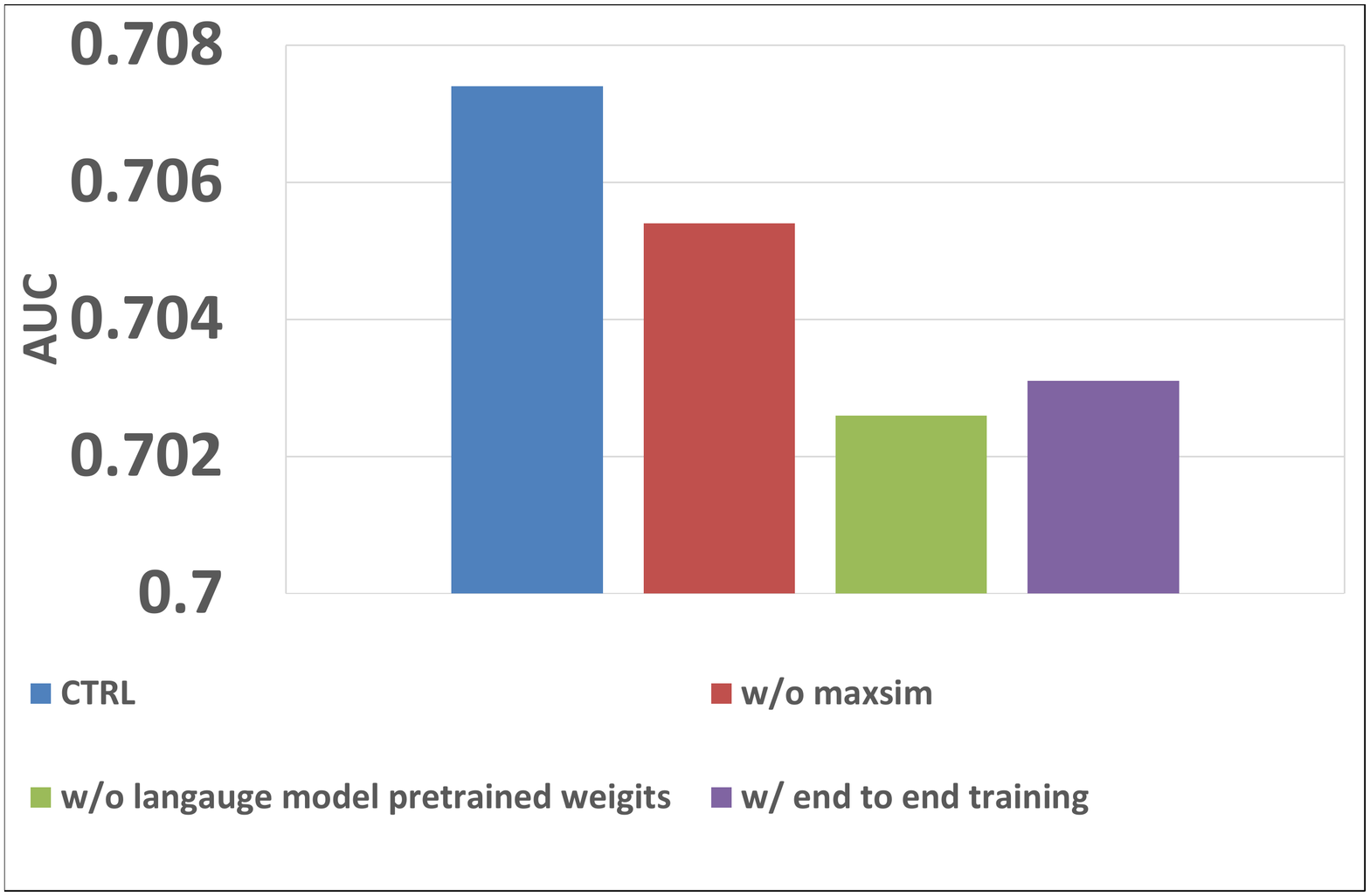}}
% 	\subfigure[\small{Alibaba}]{
% 		\label{fig:ablation:ali} 
% 		\includegraphics[height=80]{figure/ablation3.pdf}}
	\caption{\small{The results of the ablation study. }}
	\label{ablation}
\vspace{-0.8em}	
\end{figure}

In this section, we conduct ablation experiments to better understand
the importance of different components. 1)We replace the maxsim similarity with cosine similarity;  2) we remove  the pre-trained language model weights. 3) we investigate the impact of end-to-end training, which combines the two-stage process into a single stage(i.e., cross-modal knowledge alignment and CTR prediction tasks are trained together). From Figure~\ref{ablation}, we observe the following results: 1) When we remove the weights of the pre-trained language model, the loss in model performance is quite significant. This demonstrates that the primary source of improvement in the collaborative model's performance is attributed to the world knowledge and semantic modeling capabilities of the language model, rather than solely due to contrastive learning. 2) After replacing cosine similarity with maxsim similarity, there is a degradation in the model performance. This indicates that fine-grained alignment facilitates the collaborative model in learning semantic representations. 3) We observe that the performance of end-to-end training is inferior to the pre-training and fine-tuning paradigm of CTRL. We conjecture that this may be due to the multi-objective setting in end to end training paradigm, which hampers the performance of the collaborative model on the 
CTR prediction task.

\subsubsection{Prompt Analysis}
\label{sec:ablation_prompt}
%In the earlier results we presented, we only employed one type of prompt. However, previous research~\cite{p5,m6-rec} points out that various prompts can impact the effect of language modeling differently. In this subsection, we explore the effect of different prompts on the results. 
In this subsection, we explore the impact of different prompt construction methods on training CTRL. We believe that this exploration will inspire future work on how to better construct prompts. Below are several rules for constructing prompts: 1) Transform user and item features into natural language text that can be easily understood; 2) Remove auxiliary text descriptions and connect feature fields and values with ``\textbf{-}" directly; 3) Remove the feature fields and transform all the feature values into a single phrase; 4) Mask the feature fields with a meaningless unified word ``Field''; 5) Replace the separator ``\textbf{-}"  with separator ``\textbf{:}".

\begin{figure}[htbp]
\vspace{-2.0em}
	\centering
	\setlength{\belowcaptionskip}{-0.3cm}
	\setlength{\abovecaptionskip}{0cm}
	\subfigure[AUC of different prompt]{
		\includegraphics[scale=0.14]{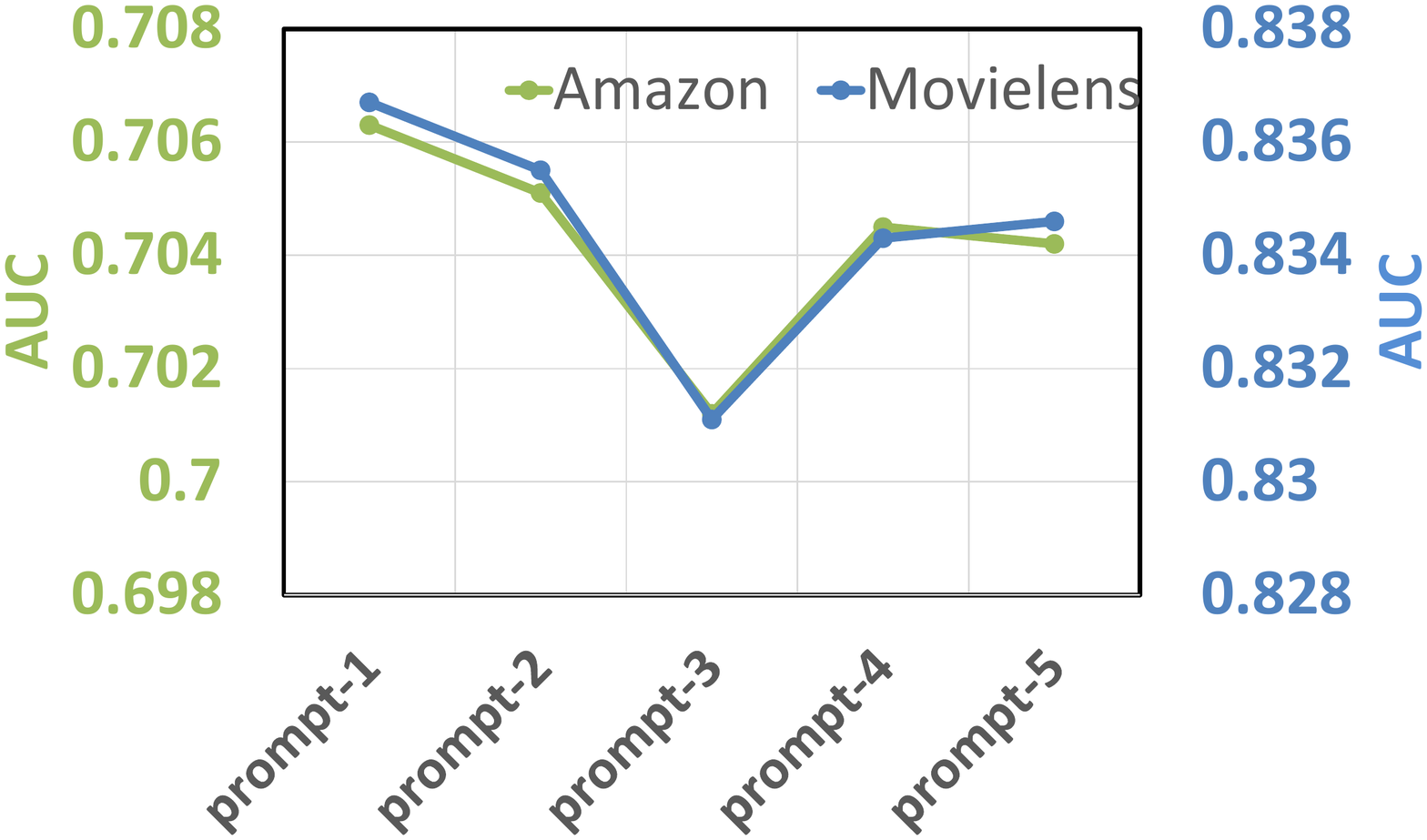}}
	\hspace{-2em}
	\subfigure[Logloss of different prompt]{
		\includegraphics[scale=0.14]{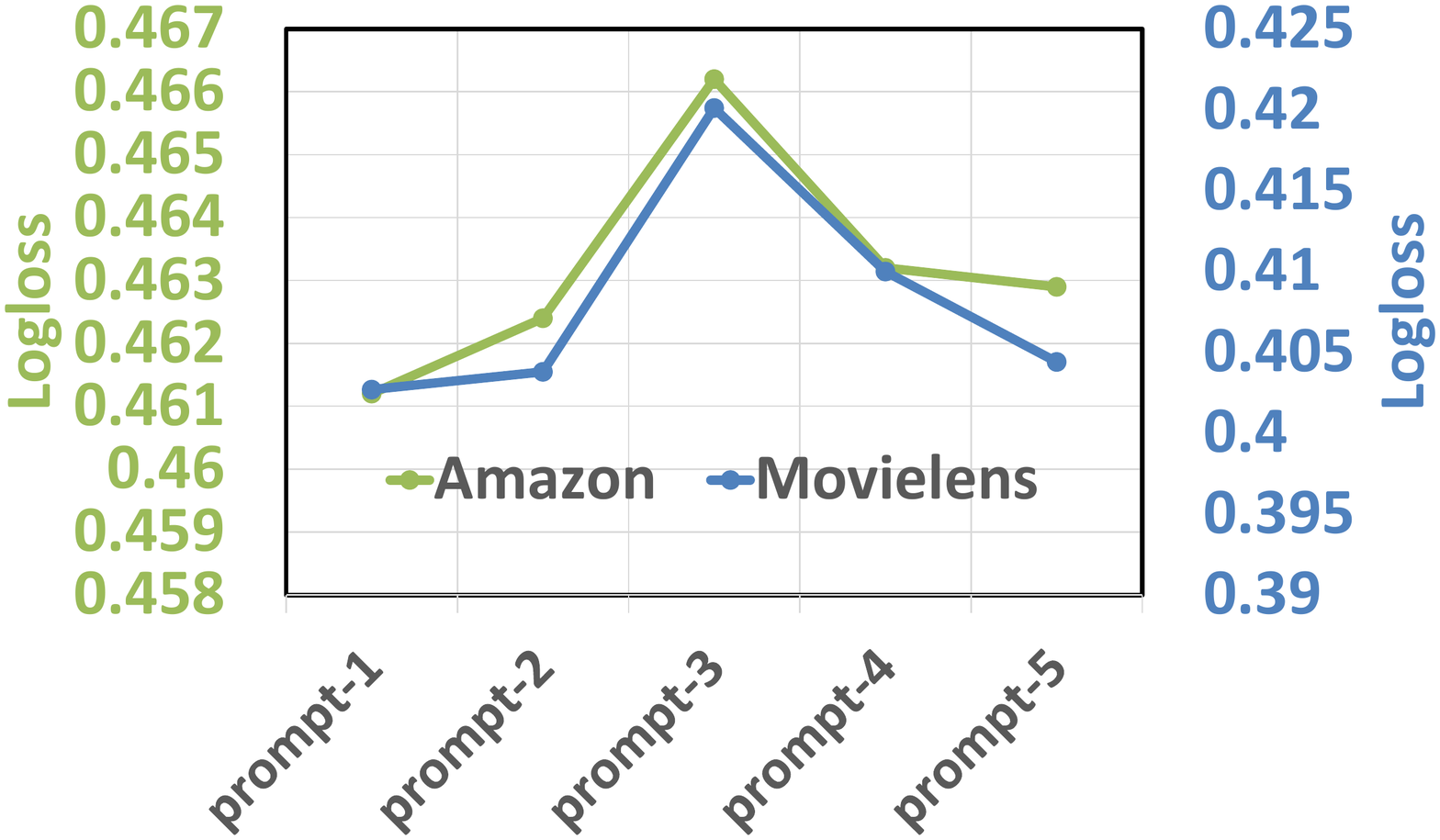}}
% 	\subfigure[\small{Alibaba}]{
% 		\label{fig:ablation:ali} 
% 		\includegraphics[height=80]{figure/ablation3.pdf}}
	\caption{\small{Performance in terms of different prompts. }}
	\label{prompt}
\end{figure}

We pre-train CTRL on these prompts and then fine-tune the CTR prediction task with the collaborative model, whose results are shown in Figure~\ref{prompt}. From Figure~\ref{prompt}, we can obtain the following observations: 1) Prompt-1 performs significantly better than all prompts, which indicates that constructing prompts in the form of natural language is beneficial for modeling. 2) The performance of Prompt-3 is weaker than Prompt-2, which confirms the importance of semantic information of feature fields, the lack of which will degrade the performance of the model remarkably. Meanwhile, the performance of Prompt-3 is weaker than Prompt-4, indicating that prompt with rules is stronger than prompt without rules. 3) The performance of Prompt-2 and Prompt-5 are similar, suggesting that the difference of connectives between feature field and feature value has little effect on the performance.
% (4) Prompt-6 with disordered order and Prompt-1 with normal order achieve close results, which verifies the consistency of the language model's 
% attention mechanism, i.e., each token is focused on all tokens.
Based on these findings, we can identify the following characteristics of designing a good prompt: 1) including feature fields such as age, gender, etc.; 2) having fluent and grammatically correct sentences and containing as much semantic information as possible.

% \cb{the conclusion is not true}
% From the above results, it is obvious that semantic information is important for prompts, and when we remove semantic information from prompts, the performance in downstream tasks decreases significantly. Therefore, we draw the conclusion in this problem of RQ3: CTRL indeed learns the semantic information from the language model as well as the semantic information is highly significant for the CTRL model.

\vspace{-0.2cm}
\section{Application in Industry System}
\vspace{-0.5em}
\begin{figure}[htbp]
\centering
\includegraphics[scale=0.32]{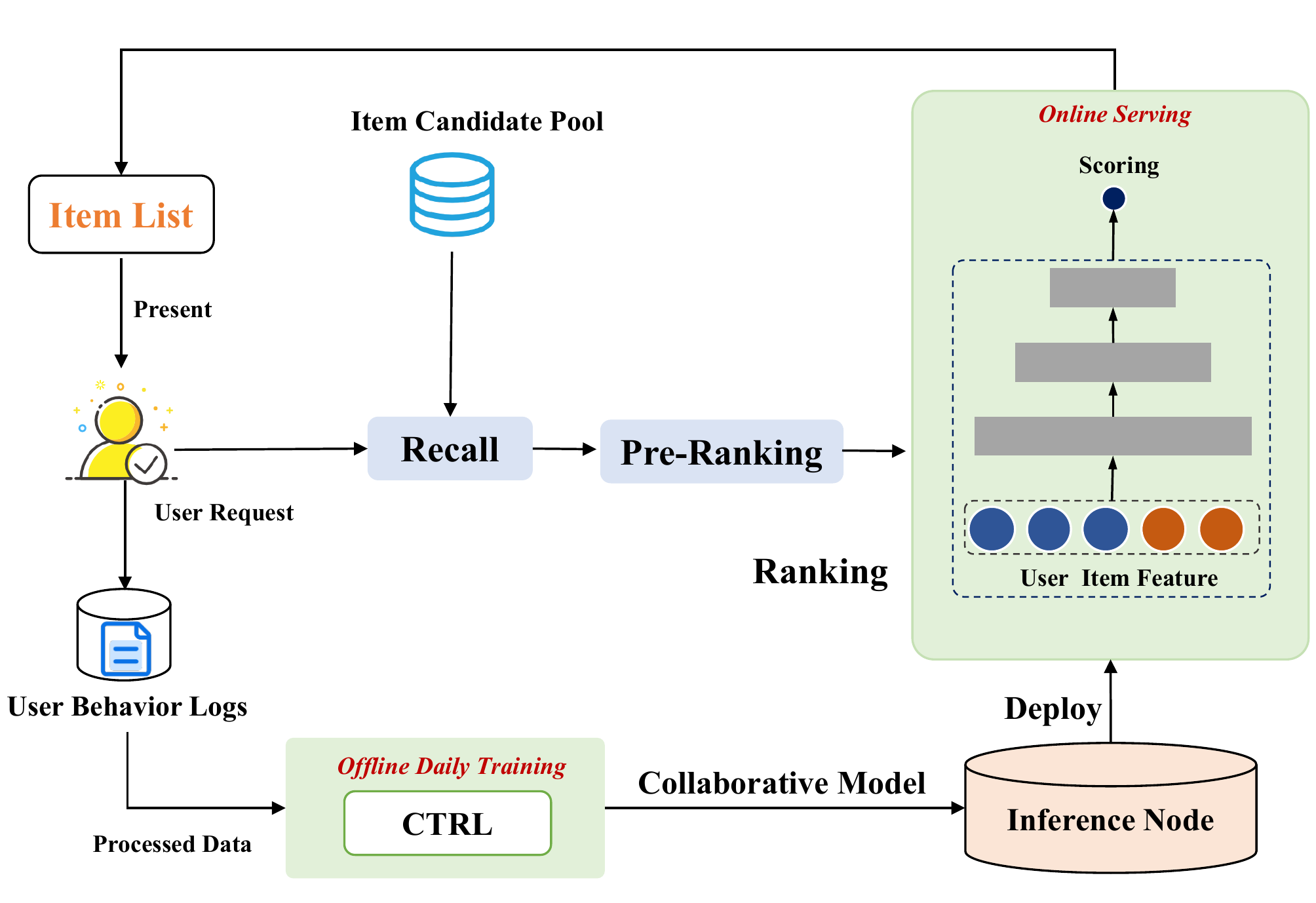}
\caption{\small{ Online workflow of CTRL.}}
\label{motivation}
\vspace{-1.5em}

\end{figure}

%介绍模型在welink私有环境的表现
\subsection{Deploying  Details of CTRL Online}
In this section, we deploy CTRL in a Huawei large-scale industrial system to verify its effectiveness. 
During the training, we collected and sampled seven days of user behavior data from Huawei large-scale recommendation platform, where millions of user logs are generated daily.
More than \textbf{30} distinct features are used, including user profile features (e.g., department), user behavior features (e.g., list of items clicked by the user), item original features (e.g., item title), and statistical features (e.g., the number of clicks on the item), as well as contextual features (e.g., time). In the first stage of the training, we only train for one epoch. In the second stage, we train for five epochs. Together, this totals to approximately five hours. This relatively short training time ensures that we are able to update the model on a daily basis. In the end, we deploy the collaborative model in CTRL at the ranking stage.
\subsection{Offline and Online Performance}
We compare the CTRL model (backbone AutoInt and RoBERTa) with the SOTA models. 
% For the semantic models, we choose CTR-BERT and P5, while for the collaborative models, we choose DeepFM, AutoInt, and DCN. 
% It is noteworthy that the industrial system under consideration comprises more than \textbf{30} distinct features.
The offline performance results are presented in Table~\ref{table:huawei}. It is evident that CTRL outperforms the baseline models significantly in terms of AUC and Logloss, thereby demonstrating its superior performance.
By incorporating the modeling capabilities of both the semantic and collaborative models, CTRL achieves a significant performance improvement over both collaborative models and semantic models.  Moreover, according to the results in Table~\ref{runtime}, CTRL would not increase any serving latency compared to the backbone collaborative model, which is an industrial-friendly framework with high accuracy and low inference latency. During the online A/B testing for seven days, we obtained a \textbf{5\%} gain of CTR compared with the base model. CTRL has now been deployed in online services, catering to tens of millions of HuaWei users.
\begin{table}[htbp]
\vspace{-1.0em}
\caption{\small{Huawei recommender system performance comparison.}}
\resizebox{0.8\linewidth}{!}{
\begin{tabular}{@{}cccccc@{}}
\toprule
Category  & Model               & AUC                        & Logloss  & RelaImpr                   \\ \midrule
\multirow{3}{*}{Collaborative}&DeepFM  &  0.6547 &   0.1801 & 8.79\%    \\
&AutoInt & \underline{0.6586}   &  \underline{0.1713} &6.12\% \\
&DCN & 0.6558   & 0.1757 &8.02\%   \\
\midrule
\multirow{2}{*}{Semantic}&CTR-BERT& 0.6484 & 0.1923 &13.41\%     \\
&P5 & 0.6472  & 0.1974 &14.33\% \\
\midrule
\multicolumn{2}{c}{CTRL}& \textbf{0.6683}$^\star$ & \textbf{0.1606}$^\star$ & -     \\
% \multicolumn{2}{1}{RelaImpr}&8.02\% &6.67\% \\

 \bottomrule
\end{tabular}
}
\label{table:huawei}
\vspace{-1.5em}
\end{table}

% \begin{table}[!t]
% % \setlength\tabcolsep{6pt}
% \caption{\small{The performance comparison in industrial  recommender system. \cb{RelaImpr}}}
% \resizebox{0.85\linewidth}{!}{
% \begin{tabular}{@{}ccccc@{}}
% \toprule
% Category  & Model               & AUC                        & Logloss & RelaImpr                   \\ \midrule
% \multirow{3}{*}{Collaborative}&DeepFM  &  0.6547 &   0.1801    &  0\%  \\
% &AutoInt & 0.6586   &  0.1713 &  2.52\% \\
% &DCN & 0.6558   &  0.1757 &  0.71\% \\
% \midrule
% \multirow{2}{*}{Semantic}&CTR-BERT& 0.6484 & 0.1923      &  -4.07\%        \\
% &P5 & 0.5594  & 0.3274 &  -61.60\% \\
% \midrule
% \multirow{1}{*}{Col \& Sem}&CTRL& 0.6683 & 0.1696      &  8.79\%        \\

%  \bottomrule
% \end{tabular}}
% \label{table:huawei}
% \end{table}

\section{Conclusion}
In this paper, we reveal the importance of both collaborative and semantic signals for CTR prediction and present CTRL, an industrial-friendly and model-agnostic framework with high inference efficiency.
CTRL treats the tabular data and converted textual data as two modalities and leverages contrastive learning for fine-grained knowledge alignment and integration. 
Finally, the lightweight collaborative model can be deployed online for efficient serving after fine-tuned with supervised signals.
Our experiments demonstrate that CTRL outperforms state-of-the-art collaborative and semantic models while maintaining good inference efficiency. Future work includes exploring the application on other downstream tasks, such as sequence recommendation and explainable recommendation.

\bibliographystyle{ACM-Reference-Format}
\bibliography{acmart}


\begin{thebibliography}{65}

%%% the Web address.

\ifx \showCODEN    \undefined \def \showCODEN     #1{\unskip}     \fi
\ifx \showDOI      \undefined \def \showDOI       #1{#1}\fi
\ifx \showISBNx    \undefined \def \showISBNx     #1{\unskip}     \fi
\ifx \showISBNxiii \undefined \def \showISBNxiii  #1{\unskip}     \fi
\ifx \showISSN     \undefined \def \showISSN      #1{\unskip}     \fi
\ifx \showLCCN     \undefined \def \showLCCN      #1{\unskip}     \fi
\ifx \shownote     \undefined \def \shownote      #1{#1}          \fi
\ifx \showarticletitle \undefined \def \showarticletitle #1{#1}   \fi
\ifx \showURL      \undefined \def \showURL       {\relax}        \fi
% invisible to TeX
\providecommand\bibfield[2]{#2}
\providecommand\bibinfo[2]{#2}
\providecommand\natexlab[1]{#1}
\providecommand\showeprint[2][]{arXiv:#2}

\bibitem[Bao et~al\mbox{.}(2023)]%
        {tallrec}
\bibfield{author}{\bibinfo{person}{Keqin Bao}, \bibinfo{person}{Jizhi Zhang}, \bibinfo{person}{Yang Zhang}, \bibinfo{person}{Wenjie Wang}, \bibinfo{person}{Fuli Feng}, {and} \bibinfo{person}{Xiangnan He}.} \bibinfo{year}{2023}\natexlab{}.
\newblock \showarticletitle{TALLRec: An Effective and Efficient Tuning Framework to Align Large Language Model with Recommendation}.
\newblock \bibinfo{journal}{\emph{arXiv preprint arXiv:2305.00447}} (\bibinfo{year}{2023}).
\newblock


\bibitem[Brown et~al\mbox{.}(2020)]%
        {gpt-3}
\bibfield{author}{\bibinfo{person}{Tom~B. Brown}, \bibinfo{person}{Benjamin Mann}, \bibinfo{person}{Nick Ryder}, \bibinfo{person}{Melanie Subbiah}, \bibinfo{person}{Jared Kaplan}, \bibinfo{person}{Prafulla Dhariwal}, \bibinfo{person}{Arvind Neelakantan}, \bibinfo{person}{Pranav Shyam}, \bibinfo{person}{Girish Sastry}, \bibinfo{person}{Amanda Askell}, \bibinfo{person}{Sandhini Agarwal}, \bibinfo{person}{Ariel Herbert-Voss}, \bibinfo{person}{Gretchen Krueger}, \bibinfo{person}{Tom Henighan}, \bibinfo{person}{Rewon Child}, \bibinfo{person}{Aditya Ramesh}, \bibinfo{person}{Daniel~M. Ziegler}, \bibinfo{person}{Jeffrey Wu}, \bibinfo{person}{Clemens Winter}, \bibinfo{person}{Christopher Hesse}, \bibinfo{person}{Mark Chen}, \bibinfo{person}{Eric Sigler}, \bibinfo{person}{Mateusz Litwin}, \bibinfo{person}{Scott Gray}, \bibinfo{person}{Benjamin Chess}, \bibinfo{person}{Jack Clark}, \bibinfo{person}{Christopher Berner}, \bibinfo{person}{Sam McCandlish}, \bibinfo{person}{Alec Radford}, \bibinfo{person}{Ilya Sutskever},
  {and} \bibinfo{person}{Dario Amodei}.} \bibinfo{year}{2020}\natexlab{}.
\newblock \bibinfo{title}{Language Models are Few-Shot Learners}.
\newblock
\newblock
\urldef\tempurl%
\url{https://doi.org/10.48550/ARXIV.2005.14165}
\showDOI{\tempurl}


\bibitem[Chen et~al\mbox{.}(2021)]%
        {edcn}
\bibfield{author}{\bibinfo{person}{Bo Chen}, \bibinfo{person}{Yichao Wang}, \bibinfo{person}{Zhirong Liu}, \bibinfo{person}{Ruiming Tang}, \bibinfo{person}{Wei Guo}, \bibinfo{person}{Hongkun Zheng}, \bibinfo{person}{Weiwei Yao}, \bibinfo{person}{Muyu Zhang}, {and} \bibinfo{person}{Xiuqiang He}.} \bibinfo{year}{2021}\natexlab{}.
\newblock \showarticletitle{Enhancing Explicit and Implicit Feature Interactions via Information Sharing for Parallel Deep CTR Models}. In \bibinfo{booktitle}{\emph{Proceedings of the 30th ACM International Conference on Information \& Knowledge Management}}. \bibinfo{pages}{3757--3766}.
\newblock


\bibitem[Chen et~al\mbox{.}(2020)]%
        {simclr}
\bibfield{author}{\bibinfo{person}{Ting Chen}, \bibinfo{person}{Simon Kornblith}, \bibinfo{person}{Mohammad Norouzi}, {and} \bibinfo{person}{Geoffrey Hinton}.} \bibinfo{year}{2020}\natexlab{}.
\newblock \showarticletitle{A simple framework for contrastive learning of visual representations}. In \bibinfo{booktitle}{\emph{Proceedings of ICML}}. PMLR, \bibinfo{pages}{1597--1607}.
\newblock


\bibitem[Chen(2023)]%
        {palr}
\bibfield{author}{\bibinfo{person}{Zheng Chen}.} \bibinfo{year}{2023}\natexlab{}.
\newblock \showarticletitle{PALR: Personalization Aware LLMs for Recommendation}.
\newblock \bibinfo{journal}{\emph{arXiv preprint arXiv:2305.07622}} (\bibinfo{year}{2023}).
\newblock


\bibitem[Cheng et~al\mbox{.}(2016)]%
        {widedeep}
\bibfield{author}{\bibinfo{person}{Heng-Tze Cheng}, \bibinfo{person}{Levent Koc}, \bibinfo{person}{Jeremiah Harmsen}, \bibinfo{person}{Tal Shaked}, \bibinfo{person}{Tushar Chandra}, \bibinfo{person}{Hrishi Aradhye}, \bibinfo{person}{Glen Anderson}, \bibinfo{person}{Greg Corrado}, \bibinfo{person}{Wei Chai}, \bibinfo{person}{Mustafa Ispir}, {et~al\mbox{.}}} \bibinfo{year}{2016}\natexlab{}.
\newblock \showarticletitle{Wide \& deep learning for recommender systems}. In \bibinfo{booktitle}{\emph{Proceedings of the 1st workshop on deep learning for recommender systems}}. \bibinfo{pages}{7--10}.
\newblock


\bibitem[Cox(1958)]%
        {LR}
\bibfield{author}{\bibinfo{person}{David~R Cox}.} \bibinfo{year}{1958}\natexlab{}.
\newblock \showarticletitle{The regression analysis of binary sequences}.
\newblock \bibinfo{journal}{\emph{Journal of the Royal Statistical Society: Series B (Methodological)}} \bibinfo{volume}{20}, \bibinfo{number}{2} (\bibinfo{year}{1958}), \bibinfo{pages}{215--232}.
\newblock


\bibitem[Cui et~al\mbox{.}(2022)]%
        {m6-rec}
\bibfield{author}{\bibinfo{person}{Zeyu Cui}, \bibinfo{person}{Jianxin Ma}, \bibinfo{person}{Chang Zhou}, \bibinfo{person}{Jingren Zhou}, {and} \bibinfo{person}{Hongxia Yang}.} \bibinfo{year}{2022}\natexlab{}.
\newblock \bibinfo{title}{M6-Rec: Generative Pretrained Language Models are Open-Ended Recommender Systems}.
\newblock
\newblock
\urldef\tempurl%
\url{https://doi.org/10.48550/ARXIV.2205.08084}
\showDOI{\tempurl}


\bibitem[Devlin et~al\mbox{.}(2018)]%
        {BERT}
\bibfield{author}{\bibinfo{person}{Jacob Devlin}, \bibinfo{person}{Ming-Wei Chang}, \bibinfo{person}{Kenton Lee}, {and} \bibinfo{person}{Kristina Toutanova}.} \bibinfo{year}{2018}\natexlab{}.
\newblock \showarticletitle{Bert: Pre-training of deep bidirectional transformers for language understanding}.
\newblock \bibinfo{journal}{\emph{arXiv preprint arXiv:1810.04805}} (\bibinfo{year}{2018}).
\newblock


\bibitem[Du et~al\mbox{.}(2022)]%
        {glm}
\bibfield{author}{\bibinfo{person}{Zhengxiao Du}, \bibinfo{person}{Yujie Qian}, \bibinfo{person}{Xiao Liu}, \bibinfo{person}{Ming Ding}, \bibinfo{person}{Jiezhong Qiu}, \bibinfo{person}{Zhilin Yang}, {and} \bibinfo{person}{Jie Tang}.} \bibinfo{year}{2022}\natexlab{}.
\newblock \showarticletitle{GLM: General Language Model Pretraining with Autoregressive Blank Infilling}. In \bibinfo{booktitle}{\emph{Proceedings of the 60th Annual Meeting of the Association for Computational Linguistics (Volume 1: Long Papers)}}. \bibinfo{pages}{320--335}.
\newblock


\bibitem[Fuglede and Topsoe(2004)]%
        {js}
\bibfield{author}{\bibinfo{person}{Bent Fuglede} {and} \bibinfo{person}{Flemming Topsoe}.} \bibinfo{year}{2004}\natexlab{}.
\newblock \showarticletitle{Jensen-Shannon divergence and Hilbert space embedding}. In \bibinfo{booktitle}{\emph{International symposium on Information theory, 2004. ISIT 2004. Proceedings.}} IEEE, \bibinfo{pages}{31}.
\newblock


\bibitem[Gai et~al\mbox{.}(2017)]%
        {taobao}
\bibfield{author}{\bibinfo{person}{Kun Gai}, \bibinfo{person}{Xiaoqiang Zhu}, \bibinfo{person}{Han Li}, \bibinfo{person}{Kai Liu}, {and} \bibinfo{person}{Zhe Wang}.} \bibinfo{year}{2017}\natexlab{}.
\newblock \showarticletitle{Learning piece-wise linear models from large scale data for ad click prediction}.
\newblock \bibinfo{journal}{\emph{arXiv preprint arXiv:1704.05194}} (\bibinfo{year}{2017}).
\newblock


\bibitem[Gao et~al\mbox{.}(2021)]%
        {gao2021simcse}
\bibfield{author}{\bibinfo{person}{Tianyu Gao}, \bibinfo{person}{Xingcheng Yao}, {and} \bibinfo{person}{Danqi Chen}.} \bibinfo{year}{2021}\natexlab{}.
\newblock \showarticletitle{{S}im{CSE}: Simple Contrastive Learning of Sentence Embeddings}. In \bibinfo{booktitle}{\emph{Proceedings of EMNLP}}. \bibinfo{pages}{6894--6910}.
\newblock


\bibitem[Geng et~al\mbox{.}(2022)]%
        {p5}
\bibfield{author}{\bibinfo{person}{Shijie Geng}, \bibinfo{person}{Shuchang Liu}, \bibinfo{person}{Zuohui Fu}, \bibinfo{person}{Yingqiang Ge}, {and} \bibinfo{person}{Yongfeng Zhang}.} \bibinfo{year}{2022}\natexlab{}.
\newblock \showarticletitle{Recommendation as language processing (rlp): A unified pretrain, personalized prompt \& predict paradigm (p5)}. In \bibinfo{booktitle}{\emph{Proceedings of the 16th ACM Conference on Recommender Systems}}. \bibinfo{pages}{299--315}.
\newblock


\bibitem[Graepel et~al\mbox{.}(2010)]%
        {graepel2010web}
\bibfield{author}{\bibinfo{person}{Thore Graepel}, \bibinfo{person}{Joaquin~Quinonero Candela}, \bibinfo{person}{Thomas Borchert}, {and} \bibinfo{person}{Ralf Herbrich}.} \bibinfo{year}{2010}\natexlab{}.
\newblock \showarticletitle{Web-scale bayesian click-through rate prediction for sponsored search advertising in microsoft's bing search engine}. Omnipress.
\newblock


\bibitem[Guo et~al\mbox{.}(2021)]%
        {autodis}
\bibfield{author}{\bibinfo{person}{Huifeng Guo}, \bibinfo{person}{Bo Chen}, \bibinfo{person}{Ruiming Tang}, \bibinfo{person}{Weinan Zhang}, \bibinfo{person}{Zhenguo Li}, {and} \bibinfo{person}{Xiuqiang He}.} \bibinfo{year}{2021}\natexlab{}.
\newblock \showarticletitle{An embedding learning framework for numerical features in ctr prediction}. In \bibinfo{booktitle}{\emph{SIGKDD}}. \bibinfo{pages}{2910--2918}.
\newblock


\bibitem[Guo et~al\mbox{.}(2017)]%
        {guo2017deepfm}
\bibfield{author}{\bibinfo{person}{Huifeng Guo}, \bibinfo{person}{Ruiming Tang}, \bibinfo{person}{Yunming Ye}, \bibinfo{person}{Zhenguo Li}, {and} \bibinfo{person}{Xiuqiang He}.} \bibinfo{year}{2017}\natexlab{}.
\newblock \showarticletitle{DeepFM: a factorization-machine based neural network for CTR prediction}.
\newblock \bibinfo{journal}{\emph{arXiv preprint arXiv:1703.04247}} (\bibinfo{year}{2017}).
\newblock


\bibitem[Gutmann and Hyv{\"a}rinen(2010)]%
        {infonce}
\bibfield{author}{\bibinfo{person}{Michael Gutmann} {and} \bibinfo{person}{Aapo Hyv{\"a}rinen}.} \bibinfo{year}{2010}\natexlab{}.
\newblock \showarticletitle{Noise-contrastive estimation: A new estimation principle for unnormalized statistical models}. In \bibinfo{booktitle}{\emph{Proceedings of AISTATS}}. JMLR Workshop and Conference Proceedings, \bibinfo{pages}{297--304}.
\newblock


\bibitem[He et~al\mbox{.}(2016)]%
        {resnet}
\bibfield{author}{\bibinfo{person}{Kaiming He}, \bibinfo{person}{Xiangyu Zhang}, \bibinfo{person}{Shaoqing Ren}, {and} \bibinfo{person}{Jian Sun}.} \bibinfo{year}{2016}\natexlab{}.
\newblock \showarticletitle{Deep residual learning for image recognition}. In \bibinfo{booktitle}{\emph{CVPR}}. \bibinfo{pages}{770--778}.
\newblock


\bibitem[He et~al\mbox{.}(2014)]%
        {he2014practical}
\bibfield{author}{\bibinfo{person}{Xinran He}, \bibinfo{person}{Junfeng Pan}, \bibinfo{person}{Ou Jin}, \bibinfo{person}{Tianbing Xu}, \bibinfo{person}{Bo Liu}, \bibinfo{person}{Tao Xu}, \bibinfo{person}{Yanxin Shi}, \bibinfo{person}{Antoine Atallah}, \bibinfo{person}{Ralf Herbrich}, \bibinfo{person}{Stuart Bowers}, {et~al\mbox{.}}} \bibinfo{year}{2014}\natexlab{}.
\newblock \showarticletitle{Practical lessons from predicting clicks on ads at facebook}. In \bibinfo{booktitle}{\emph{Proceedings of the eighth international workshop on data mining for online advertising}}. \bibinfo{pages}{1--9}.
\newblock


\bibitem[Hoang et~al\mbox{.}(2019)]%
        {sa-2}
\bibfield{author}{\bibinfo{person}{Mickel Hoang}, \bibinfo{person}{Oskar~Alija Bihorac}, {and} \bibinfo{person}{Jacobo Rouces}.} \bibinfo{year}{2019}\natexlab{}.
\newblock \showarticletitle{Aspect-based sentiment analysis using bert}. In \bibinfo{booktitle}{\emph{Proceedings of the 22nd nordic conference on computational linguistics}}. \bibinfo{pages}{187--196}.
\newblock


\bibitem[Hou et~al\mbox{.}(2023)]%
        {gpt-rank}
\bibfield{author}{\bibinfo{person}{Yupeng Hou}, \bibinfo{person}{Junjie Zhang}, \bibinfo{person}{Zihan Lin}, \bibinfo{person}{Hongyu Lu}, \bibinfo{person}{Ruobing Xie}, \bibinfo{person}{Julian McAuley}, {and} \bibinfo{person}{Wayne~Xin Zhao}.} \bibinfo{year}{2023}\natexlab{}.
\newblock \showarticletitle{Large Language Models are Zero-Shot Rankers for Recommender Systems}.
\newblock \bibinfo{journal}{\emph{arXiv preprint arXiv:2305.08845}} (\bibinfo{year}{2023}).
\newblock


\bibitem[Hu et~al\mbox{.}(2022)]%
        {transfer-1}
\bibfield{author}{\bibinfo{person}{Zhiqiang Hu}, \bibinfo{person}{Roy Ka-Wei Lee}, \bibinfo{person}{Charu~C Aggarwal}, {and} \bibinfo{person}{Aston Zhang}.} \bibinfo{year}{2022}\natexlab{}.
\newblock \showarticletitle{Text style transfer: A review and experimental evaluation}.
\newblock \bibinfo{journal}{\emph{ACM SIGKDD Explorations Newsletter}} \bibinfo{volume}{24}, \bibinfo{number}{1} (\bibinfo{year}{2022}), \bibinfo{pages}{14--45}.
\newblock


\bibitem[Huang et~al\mbox{.}(2019a)]%
        {fibinet}
\bibfield{author}{\bibinfo{person}{Tongwen Huang}, \bibinfo{person}{Zhiqi Zhang}, {and} \bibinfo{person}{Junlin Zhang}.} \bibinfo{year}{2019}\natexlab{a}.
\newblock \showarticletitle{{FiBiNET}}. In \bibinfo{booktitle}{\emph{Proceedings of the 13th {ACM} Conference on Recommender Systems}}. \bibinfo{publisher}{{ACM}}.
\newblock
\urldef\tempurl%
\url{https://doi.org/10.1145/3298689.3347043}
\showDOI{\tempurl}


\bibitem[Huang et~al\mbox{.}(2019b)]%
        {huang2019fibinet}
\bibfield{author}{\bibinfo{person}{Tongwen Huang}, \bibinfo{person}{Zhiqi Zhang}, {and} \bibinfo{person}{Junlin Zhang}.} \bibinfo{year}{2019}\natexlab{b}.
\newblock \showarticletitle{FiBiNET: combining feature importance and bilinear feature interaction for click-through rate prediction}. In \bibinfo{booktitle}{\emph{Proceedings of the 13th ACM Conference on Recommender Systems}}. \bibinfo{pages}{169--177}.
\newblock


\bibitem[Ioffe and Szegedy(2015)]%
        {bn}
\bibfield{author}{\bibinfo{person}{Sergey Ioffe} {and} \bibinfo{person}{Christian Szegedy}.} \bibinfo{year}{2015}\natexlab{}.
\newblock \showarticletitle{Batch normalization: Accelerating deep network training by reducing internal covariate shift}. In \bibinfo{booktitle}{\emph{ICML}}. PMLR, \bibinfo{pages}{448--456}.
\newblock


\bibitem[Jiao et~al\mbox{.}(2019)]%
        {tinyBERT}
\bibfield{author}{\bibinfo{person}{Xiaoqi Jiao}, \bibinfo{person}{Yichun Yin}, \bibinfo{person}{Lifeng Shang}, \bibinfo{person}{Xin Jiang}, \bibinfo{person}{Xiao Chen}, \bibinfo{person}{Linlin Li}, \bibinfo{person}{Fang Wang}, {and} \bibinfo{person}{Qun Liu}.} \bibinfo{year}{2019}\natexlab{}.
\newblock \showarticletitle{Tinybert: Distilling bert for natural language understanding}.
\newblock \bibinfo{journal}{\emph{arXiv preprint arXiv:1909.10351}} (\bibinfo{year}{2019}).
\newblock


\bibitem[Kingma and Ba(2014)]%
        {https://doi.org/10.48550/arxiv.1412.6980}
\bibfield{author}{\bibinfo{person}{Diederik~P. Kingma} {and} \bibinfo{person}{Jimmy Ba}.} \bibinfo{year}{2014}\natexlab{}.
\newblock \bibinfo{title}{Adam: A Method for Stochastic Optimization}.
\newblock
\newblock
\urldef\tempurl%
\url{https://doi.org/10.48550/ARXIV.1412.6980}
\showDOI{\tempurl}


\bibitem[Lee and Seung(1999)]%
        {mf}
\bibfield{author}{\bibinfo{person}{Daniel~D Lee} {and} \bibinfo{person}{H~Sebastian Seung}.} \bibinfo{year}{1999}\natexlab{}.
\newblock \showarticletitle{Learning the parts of objects by non-negative matrix factorization}.
\newblock \bibinfo{journal}{\emph{Nature}} \bibinfo{volume}{401}, \bibinfo{number}{6755} (\bibinfo{year}{1999}), \bibinfo{pages}{788--791}.
\newblock


\bibitem[Li et~al\mbox{.}(2022a)]%
        {inttower}
\bibfield{author}{\bibinfo{person}{Xiangyang Li}, \bibinfo{person}{Bo Chen}, \bibinfo{person}{HuiFeng Guo}, \bibinfo{person}{Jingjie Li}, \bibinfo{person}{Chenxu Zhu}, \bibinfo{person}{Xiang Long}, \bibinfo{person}{Sujian Li}, \bibinfo{person}{Yichao Wang}, \bibinfo{person}{Wei Guo}, \bibinfo{person}{Longxia Mao}, {et~al\mbox{.}}} \bibinfo{year}{2022}\natexlab{a}.
\newblock \showarticletitle{IntTower: the Next Generation of Two-Tower Model for Pre-Ranking System}. In \bibinfo{booktitle}{\emph{CIKM}}. \bibinfo{pages}{3292--3301}.
\newblock


\bibitem[Li et~al\mbox{.}(2022b)]%
        {transfer-2}
\bibfield{author}{\bibinfo{person}{Xiangyang Li}, \bibinfo{person}{Xiang Long}, \bibinfo{person}{Yu Xia}, {and} \bibinfo{person}{Sujian Li}.} \bibinfo{year}{2022}\natexlab{b}.
\newblock \showarticletitle{Low Resource Style Transfer via Domain Adaptive Meta Learning}.
\newblock \bibinfo{journal}{\emph{arXiv preprint arXiv:2205.12475}} (\bibinfo{year}{2022}).
\newblock


\bibitem[Li et~al\mbox{.}(2021)]%
        {text-2}
\bibfield{author}{\bibinfo{person}{Xiangyang Li}, \bibinfo{person}{Yu Xia}, \bibinfo{person}{Xiang Long}, \bibinfo{person}{Zheng Li}, {and} \bibinfo{person}{Sujian Li}.} \bibinfo{year}{2021}\natexlab{}.
\newblock \showarticletitle{Exploring text-transformers in aaai 2021 shared task: Covid-19 fake news detection in english}. In \bibinfo{booktitle}{\emph{Combating Online Hostile Posts in Regional Languages during Emergency Situation: First International Workshop, CONSTRAINT 2021, Collocated with AAAI 2021, Virtual Event, February 8, 2021, Revised Selected Papers 1}}. Springer, \bibinfo{pages}{106--115}.
\newblock


\bibitem[Lian et~al\mbox{.}(2018)]%
        {xdeepfm}
\bibfield{author}{\bibinfo{person}{Jianxun Lian}, \bibinfo{person}{Xiaohuan Zhou}, \bibinfo{person}{Fuzheng Zhang}, \bibinfo{person}{Zhongxia Chen}, \bibinfo{person}{Xing Xie}, {and} \bibinfo{person}{Guangzhong Sun}.} \bibinfo{year}{2018}\natexlab{}.
\newblock \showarticletitle{xdeepfm: Combining explicit and implicit feature interactions for recommender systems}. In \bibinfo{booktitle}{\emph{SIGKDD}}. \bibinfo{pages}{1754--1763}.
\newblock


\bibitem[Liu et~al\mbox{.}(2019b)]%
        {fgcnn}
\bibfield{author}{\bibinfo{person}{Bin Liu}, \bibinfo{person}{Ruiming Tang}, \bibinfo{person}{Yingzhi Chen}, \bibinfo{person}{Jinkai Yu}, \bibinfo{person}{Huifeng Guo}, {and} \bibinfo{person}{Yuzhou Zhang}.} \bibinfo{year}{2019}\natexlab{b}.
\newblock \showarticletitle{Feature generation by convolutional neural network for click-through rate prediction}. In \bibinfo{booktitle}{\emph{The World Wide Web Conference}}. \bibinfo{pages}{1119--1129}.
\newblock


\bibitem[Liu et~al\mbox{.}(2022)]%
        {ptab}
\bibfield{author}{\bibinfo{person}{Guang Liu}, \bibinfo{person}{Jie Yang}, {and} \bibinfo{person}{Ledell Wu}.} \bibinfo{year}{2022}\natexlab{}.
\newblock \showarticletitle{PTab: Using the Pre-trained Language Model for Modeling Tabular Data}.
\newblock \bibinfo{journal}{\emph{arXiv preprint arXiv:2209.08060}} (\bibinfo{year}{2022}).
\newblock


\bibitem[Liu et~al\mbox{.}(2023)]%
        {liu2023chatgpt}
\bibfield{author}{\bibinfo{person}{Junling Liu}, \bibinfo{person}{Chao Liu}, \bibinfo{person}{Renjie Lv}, \bibinfo{person}{Kang Zhou}, {and} \bibinfo{person}{Yan Zhang}.} \bibinfo{year}{2023}\natexlab{}.
\newblock \showarticletitle{Is ChatGPT a Good Recommender? A Preliminary Study}.
\newblock \bibinfo{journal}{\emph{arXiv preprint arXiv:2304.10149}} (\bibinfo{year}{2023}).
\newblock


\bibitem[Liu et~al\mbox{.}(2019a)]%
        {roBERTa}
\bibfield{author}{\bibinfo{person}{Yinhan Liu}, \bibinfo{person}{Myle Ott}, \bibinfo{person}{Naman Goyal}, \bibinfo{person}{Jingfei Du}, \bibinfo{person}{Mandar Joshi}, \bibinfo{person}{Danqi Chen}, \bibinfo{person}{Omer Levy}, \bibinfo{person}{Mike Lewis}, \bibinfo{person}{Luke Zettlemoyer}, {and} \bibinfo{person}{Veselin Stoyanov}.} \bibinfo{year}{2019}\natexlab{a}.
\newblock \showarticletitle{Roberta: A robustly optimized bert pretraining approach}.
\newblock \bibinfo{journal}{\emph{arXiv preprint arXiv:1907.11692}} (\bibinfo{year}{2019}).
\newblock


\bibitem[Loshchilov and Hutter(2017)]%
        {adamw}
\bibfield{author}{\bibinfo{person}{Ilya Loshchilov} {and} \bibinfo{person}{Frank Hutter}.} \bibinfo{year}{2017}\natexlab{}.
\newblock \showarticletitle{Decoupled weight decay regularization}.
\newblock \bibinfo{journal}{\emph{arXiv preprint arXiv:1711.05101}} (\bibinfo{year}{2017}).
\newblock


\bibitem[Lu et~al\mbox{.}(2020)]%
        {lu2020meta}
\bibfield{author}{\bibinfo{person}{Yuanfu Lu}, \bibinfo{person}{Yuan Fang}, {and} \bibinfo{person}{Chuan Shi}.} \bibinfo{year}{2020}\natexlab{}.
\newblock \showarticletitle{Meta-learning on heterogeneous information networks for cold-start recommendation}. In \bibinfo{booktitle}{\emph{Proceedings of the 26th ACM SIGKDD International Conference on Knowledge Discovery \& Data Mining}}. \bibinfo{pages}{1563--1573}.
\newblock


\bibitem[McMahan et~al\mbox{.}(2013)]%
        {mcmahan2013ad}
\bibfield{author}{\bibinfo{person}{H~Brendan McMahan}, \bibinfo{person}{Gary Holt}, \bibinfo{person}{David Sculley}, \bibinfo{person}{Michael Young}, \bibinfo{person}{Dietmar Ebner}, \bibinfo{person}{Julian Grady}, \bibinfo{person}{Lan Nie}, \bibinfo{person}{Todd Phillips}, \bibinfo{person}{Eugene Davydov}, \bibinfo{person}{Daniel Golovin}, {et~al\mbox{.}}} \bibinfo{year}{2013}\natexlab{}.
\newblock \showarticletitle{Ad click prediction: a view from the trenches}. In \bibinfo{booktitle}{\emph{SIGKDD}}. \bibinfo{pages}{1222--1230}.
\newblock


\bibitem[Miro{\'n}czuk and Protasiewicz(2018)]%
        {text-1}
\bibfield{author}{\bibinfo{person}{Marcin~Micha{\l} Miro{\'n}czuk} {and} \bibinfo{person}{Jaros{\l}aw Protasiewicz}.} \bibinfo{year}{2018}\natexlab{}.
\newblock \showarticletitle{A recent overview of the state-of-the-art elements of text classification}.
\newblock \bibinfo{journal}{\emph{Expert Systems with Applications}}  \bibinfo{volume}{106} (\bibinfo{year}{2018}), \bibinfo{pages}{36--54}.
\newblock


\bibitem[Muhamed et~al\mbox{.}(2021)]%
        {ctr-BERT}
\bibfield{author}{\bibinfo{person}{Aashiq Muhamed}, \bibinfo{person}{Iman Keivanloo}, \bibinfo{person}{Sujan Perera}, \bibinfo{person}{James Mracek}, \bibinfo{person}{Yi Xu}, \bibinfo{person}{Qingjun Cui}, \bibinfo{person}{Santosh Rajagopalan}, \bibinfo{person}{Belinda Zeng}, {and} \bibinfo{person}{Trishul Chilimbi}.} \bibinfo{year}{2021}\natexlab{}.
\newblock \showarticletitle{CTR-BERT: Cost-effective knowledge distillation for billion-parameter teacher models}. In \bibinfo{booktitle}{\emph{NeurIPS Efficient Natural Language and Speech Processing Workshop}}.
\newblock


\bibitem[Ni et~al\mbox{.}(2019)]%
        {amzon}
\bibfield{author}{\bibinfo{person}{Jianmo Ni}, \bibinfo{person}{Jiacheng Li}, {and} \bibinfo{person}{Julian McAuley}.} \bibinfo{year}{2019}\natexlab{}.
\newblock \showarticletitle{Justifying recommendations using distantly-labeled reviews and fine-grained aspects}. In \bibinfo{booktitle}{\emph{EMNLP-IJCNLP}}. \bibinfo{pages}{188--197}.
\newblock


\bibitem[Ouyang et~al\mbox{.}(2022)]%
        {instruct-gpt}
\bibfield{author}{\bibinfo{person}{Long Ouyang}, \bibinfo{person}{Jeff Wu}, \bibinfo{person}{Xu Jiang}, \bibinfo{person}{Diogo Almeida}, \bibinfo{person}{Carroll~L Wainwright}, \bibinfo{person}{Pamela Mishkin}, \bibinfo{person}{Chong Zhang}, \bibinfo{person}{Sandhini Agarwal}, \bibinfo{person}{Katarina Slama}, \bibinfo{person}{Alex Ray}, {et~al\mbox{.}}} \bibinfo{year}{2022}\natexlab{}.
\newblock \showarticletitle{Training language models to follow instructions with human feedback}.
\newblock \bibinfo{journal}{\emph{arXiv preprint arXiv:2203.02155}} (\bibinfo{year}{2022}).
\newblock


\bibitem[Qu et~al\mbox{.}(2016)]%
        {pnn}
\bibfield{author}{\bibinfo{person}{Yanru Qu}, \bibinfo{person}{Han Cai}, \bibinfo{person}{Kan Ren}, \bibinfo{person}{Weinan Zhang}, \bibinfo{person}{Yong Yu}, \bibinfo{person}{Ying Wen}, {and} \bibinfo{person}{Jun Wang}.} \bibinfo{year}{2016}\natexlab{}.
\newblock \showarticletitle{Product-based neural networks for user response prediction}. In \bibinfo{booktitle}{\emph{ICDM}}. IEEE, \bibinfo{pages}{1149--1154}.
\newblock


\bibitem[Radford et~al\mbox{.}(2021)]%
        {clip}
\bibfield{author}{\bibinfo{person}{Alec Radford}, \bibinfo{person}{Jong~Wook Kim}, \bibinfo{person}{Chris Hallacy}, \bibinfo{person}{Aditya Ramesh}, \bibinfo{person}{Gabriel Goh}, \bibinfo{person}{Sandhini Agarwal}, \bibinfo{person}{Girish Sastry}, \bibinfo{person}{Amanda Askell}, \bibinfo{person}{Pamela Mishkin}, \bibinfo{person}{Jack Clark}, {et~al\mbox{.}}} \bibinfo{year}{2021}\natexlab{}.
\newblock \showarticletitle{Learning transferable visual models from natural language supervision}. In \bibinfo{booktitle}{\emph{International conference on machine learning}}. PMLR, \bibinfo{pages}{8748--8763}.
\newblock


\bibitem[Raffel et~al\mbox{.}(2019)]%
        {t5}
\bibfield{author}{\bibinfo{person}{Colin Raffel}, \bibinfo{person}{Noam Shazeer}, \bibinfo{person}{Adam Roberts}, \bibinfo{person}{Katherine Lee}, \bibinfo{person}{Sharan Narang}, \bibinfo{person}{Michael Matena}, \bibinfo{person}{Yanqi Zhou}, \bibinfo{person}{Wei Li}, {and} \bibinfo{person}{Peter~J. Liu}.} \bibinfo{year}{2019}\natexlab{}.
\newblock \bibinfo{title}{Exploring the Limits of Transfer Learning with a Unified Text-to-Text Transformer}.
\newblock
\newblock
\urldef\tempurl%
\url{https://doi.org/10.48550/ARXIV.1910.10683}
\showDOI{\tempurl}


\bibitem[Rendle(2010)]%
        {FM}
\bibfield{author}{\bibinfo{person}{Steffen Rendle}.} \bibinfo{year}{2010}\natexlab{}.
\newblock \showarticletitle{Factorization machines}. In \bibinfo{booktitle}{\emph{2010 IEEE International conference on data mining}}. IEEE, \bibinfo{pages}{995--1000}.
\newblock


\bibitem[Song et~al\mbox{.}(2019)]%
        {autoint}
\bibfield{author}{\bibinfo{person}{Weiping Song}, \bibinfo{person}{Chence Shi}, \bibinfo{person}{Zhiping Xiao}, \bibinfo{person}{Zhijian Duan}, \bibinfo{person}{Yewen Xu}, \bibinfo{person}{Ming Zhang}, {and} \bibinfo{person}{Jian Tang}.} \bibinfo{year}{2019}\natexlab{}.
\newblock \showarticletitle{Autoint: Automatic feature interaction learning via self-attentive neural networks}. In \bibinfo{booktitle}{\emph{CIKM}}. \bibinfo{pages}{1161--1170}.
\newblock


\bibitem[Srivastava et~al\mbox{.}(2014)]%
        {dropout}
\bibfield{author}{\bibinfo{person}{Nitish Srivastava}, \bibinfo{person}{Geoffrey Hinton}, \bibinfo{person}{Alex Krizhevsky}, \bibinfo{person}{Ilya Sutskever}, {and} \bibinfo{person}{Ruslan Salakhutdinov}.} \bibinfo{year}{2014}\natexlab{}.
\newblock \showarticletitle{Dropout: a simple way to prevent neural networks from overfitting}.
\newblock \bibinfo{journal}{\emph{The journal of machine learning research}} \bibinfo{volume}{15}, \bibinfo{number}{1} (\bibinfo{year}{2014}), \bibinfo{pages}{1929--1958}.
\newblock


\bibitem[Sun et~al\mbox{.}(2023)]%
        {gpt-search}
\bibfield{author}{\bibinfo{person}{Weiwei Sun}, \bibinfo{person}{Lingyong Yan}, \bibinfo{person}{Xinyu Ma}, \bibinfo{person}{Pengjie Ren}, \bibinfo{person}{Dawei Yin}, {and} \bibinfo{person}{Zhaochun Ren}.} \bibinfo{year}{2023}\natexlab{}.
\newblock \showarticletitle{Is ChatGPT Good at Search? Investigating Large Language Models as Re-Ranking Agent}.
\newblock \bibinfo{journal}{\emph{arXiv preprint arXiv:2304.09542}} (\bibinfo{year}{2023}).
\newblock


\bibitem[Van~der Maaten and Hinton(2008)]%
        {t-SNE}
\bibfield{author}{\bibinfo{person}{Laurens Van~der Maaten} {and} \bibinfo{person}{Geoffrey Hinton}.} \bibinfo{year}{2008}\natexlab{}.
\newblock \showarticletitle{Visualizing data using t-SNE.}
\newblock \bibinfo{journal}{\emph{Journal of machine learning research}} \bibinfo{volume}{9}, \bibinfo{number}{11} (\bibinfo{year}{2008}).
\newblock


\bibitem[Wang et~al\mbox{.}(2020)]%
        {superglue}
\bibfield{author}{\bibinfo{person}{Alex Wang}, \bibinfo{person}{Yada Pruksachatkun}, \bibinfo{person}{Nikita Nangia}, \bibinfo{person}{Amanpreet Singh}, \bibinfo{person}{Julian Michael}, \bibinfo{person}{Felix Hill}, \bibinfo{person}{Omer Levy}, {and} \bibinfo{person}{Samuel~R. Bowman}.} \bibinfo{year}{2020}\natexlab{}.
\newblock \bibinfo{title}{SuperGLUE: A Stickier Benchmark for General-Purpose Language Understanding Systems}.
\newblock
\newblock
\showeprint[arxiv]{1905.00537}~[cs.CL]


\bibitem[Wang et~al\mbox{.}(2017)]%
        {dcn}
\bibfield{author}{\bibinfo{person}{Ruoxi Wang}, \bibinfo{person}{Bin Fu}, \bibinfo{person}{Gang Fu}, {and} \bibinfo{person}{Mingliang Wang}.} \bibinfo{year}{2017}\natexlab{}.
\newblock \showarticletitle{Deep \& cross network for ad click predictions}.
\newblock In \bibinfo{booktitle}{\emph{ADKDD}}. \bibinfo{pages}{1--7}.
\newblock


\bibitem[Xin et~al\mbox{.}(2019)]%
        {cfm}
\bibfield{author}{\bibinfo{person}{Xin Xin}, \bibinfo{person}{Bo Chen}, \bibinfo{person}{Xiangnan He}, \bibinfo{person}{Dong Wang}, \bibinfo{person}{Yue Ding}, {and} \bibinfo{person}{Joemon~M Jose}.} \bibinfo{year}{2019}\natexlab{}.
\newblock \showarticletitle{CFM: Convolutional Factorization Machines for Context-Aware Recommendation.}. In \bibinfo{booktitle}{\emph{IJCAI}}, Vol.~\bibinfo{volume}{19}. \bibinfo{pages}{3926--3932}.
\newblock


\bibitem[Xu et~al\mbox{.}(2019)]%
        {sa-1}
\bibfield{author}{\bibinfo{person}{Hu Xu}, \bibinfo{person}{Bing Liu}, \bibinfo{person}{Lei Shu}, {and} \bibinfo{person}{Philip~S Yu}.} \bibinfo{year}{2019}\natexlab{}.
\newblock \showarticletitle{BERT post-training for review reading comprehension and aspect-based sentiment analysis}.
\newblock \bibinfo{journal}{\emph{arXiv preprint arXiv:1904.02232}} (\bibinfo{year}{2019}).
\newblock


\bibitem[Yao et~al\mbox{.}(2021)]%
        {filip}
\bibfield{author}{\bibinfo{person}{Lewei Yao}, \bibinfo{person}{Runhui Huang}, \bibinfo{person}{Lu Hou}, \bibinfo{person}{Guansong Lu}, \bibinfo{person}{Minzhe Niu}, \bibinfo{person}{Hang Xu}, \bibinfo{person}{Xiaodan Liang}, \bibinfo{person}{Zhenguo Li}, \bibinfo{person}{Xin Jiang}, {and} \bibinfo{person}{Chunjing Xu}.} \bibinfo{year}{2021}\natexlab{}.
\newblock \showarticletitle{FILIP: fine-grained interactive language-image pre-training}.
\newblock \bibinfo{journal}{\emph{arXiv preprint arXiv:2111.07783}} (\bibinfo{year}{2021}).
\newblock


\bibitem[Yu et~al\mbox{.}(2021)]%
        {dat}
\bibfield{author}{\bibinfo{person}{Yantao Yu}, \bibinfo{person}{Weipeng Wang}, \bibinfo{person}{Zhoutian Feng}, {and} \bibinfo{person}{Daiyue Xue}.} \bibinfo{year}{2021}\natexlab{}.
\newblock \showarticletitle{A Dual Augmented Two-tower Model for Online Large-scale Recommendation}.
\newblock  (\bibinfo{year}{2021}).
\newblock


\bibitem[Yu et~al\mbox{.}(2019)]%
        {yu2019adaptive}
\bibfield{author}{\bibinfo{person}{Zeping Yu}, \bibinfo{person}{Jianxun Lian}, \bibinfo{person}{Ahmad Mahmoody}, \bibinfo{person}{Gongshen Liu}, {and} \bibinfo{person}{Xing Xie}.} \bibinfo{year}{2019}\natexlab{}.
\newblock \showarticletitle{Adaptive User Modeling with Long and Short-Term Preferences for Personalized Recommendation.}. In \bibinfo{booktitle}{\emph{IJCAI}}. \bibinfo{pages}{4213--4219}.
\newblock


\bibitem[Zhang et~al\mbox{.}(2023a)]%
        {llm-fair}
\bibfield{author}{\bibinfo{person}{Jizhi Zhang}, \bibinfo{person}{Keqin Bao}, \bibinfo{person}{Yang Zhang}, \bibinfo{person}{Wenjie Wang}, \bibinfo{person}{Fuli Feng}, {and} \bibinfo{person}{Xiangnan He}.} \bibinfo{year}{2023}\natexlab{a}.
\newblock \showarticletitle{Is ChatGPT Fair for Recommendation? Evaluating Fairness in Large Language Model Recommendation}.
\newblock \bibinfo{journal}{\emph{arXiv preprint arXiv:2305.07609}} (\bibinfo{year}{2023}).
\newblock


\bibitem[Zhang et~al\mbox{.}(2023b)]%
        {InstructRec}
\bibfield{author}{\bibinfo{person}{Junjie Zhang}, \bibinfo{person}{Ruobing Xie}, \bibinfo{person}{Yupeng Hou}, \bibinfo{person}{Wayne~Xin Zhao}, \bibinfo{person}{Leyu Lin}, {and} \bibinfo{person}{Ji-Rong Wen}.} \bibinfo{year}{2023}\natexlab{b}.
\newblock \showarticletitle{Recommendation as instruction following: A large language model empowered recommendation approach}.
\newblock \bibinfo{journal}{\emph{arXiv preprint arXiv:2305.07001}} (\bibinfo{year}{2023}).
\newblock


\bibitem[Zhang et~al\mbox{.}(2016)]%
        {zhang2016deep}
\bibfield{author}{\bibinfo{person}{Weinan Zhang}, \bibinfo{person}{Tianming Du}, {and} \bibinfo{person}{Jun Wang}.} \bibinfo{year}{2016}\natexlab{}.
\newblock \showarticletitle{Deep learning over multi-field categorical data}. In \bibinfo{booktitle}{\emph{ECIR}}. Springer, \bibinfo{pages}{45--57}.
\newblock


\bibitem[Zhang et~al\mbox{.}(2014)]%
        {zhang2014optimal}
\bibfield{author}{\bibinfo{person}{Weinan Zhang}, \bibinfo{person}{Shuai Yuan}, {and} \bibinfo{person}{Jun Wang}.} \bibinfo{year}{2014}\natexlab{}.
\newblock \showarticletitle{Optimal real-time bidding for display advertising}. In \bibinfo{booktitle}{\emph{Proceedings of the 20th ACM SIGKDD international conference on Knowledge discovery and data mining}}. \bibinfo{pages}{1077--1086}.
\newblock


\bibitem[Zhang et~al\mbox{.}(2021)]%
        {zhang2021language}
\bibfield{author}{\bibinfo{person}{Yuhui Zhang}, \bibinfo{person}{Hao Ding}, \bibinfo{person}{Zeren Shui}, \bibinfo{person}{Yifei Ma}, \bibinfo{person}{James Zou}, \bibinfo{person}{Anoop Deoras}, {and} \bibinfo{person}{Hao Wang}.} \bibinfo{year}{2021}\natexlab{}.
\newblock \showarticletitle{Language models as recommender systems: Evaluations and limitations}.
\newblock  (\bibinfo{year}{2021}).
\newblock


\bibitem[Zhou et~al\mbox{.}(2018)]%
        {din}
\bibfield{author}{\bibinfo{person}{Guorui Zhou}, \bibinfo{person}{Xiaoqiang Zhu}, \bibinfo{person}{Chenru Song}, \bibinfo{person}{Ying Fan}, \bibinfo{person}{Han Zhu}, \bibinfo{person}{Xiao Ma}, \bibinfo{person}{Yanghui Yan}, \bibinfo{person}{Junqi Jin}, \bibinfo{person}{Han Li}, {and} \bibinfo{person}{Kun Gai}.} \bibinfo{year}{2018}\natexlab{}.
\newblock \showarticletitle{Deep interest network for click-through rate prediction}. In \bibinfo{booktitle}{\emph{SIGKDD}}. \bibinfo{pages}{1059--1068}.
\newblock


\end{thebibliography}

\appendix
\section{Experimental Setting}
\label{appendix_setting}
\subsection{Datasets and Evaluation Metrics}
\textbf{MovieLens  Dataset\footnote{https://grouplens.org/datasets/MovieLens/1m/}} is a movie recommendation dataset and following previous work~\cite{autoint}, we consider samples with ratings less than 3 as negative, samples with scores greater than 3 as positive, and remove neutral samples, i.e., rating equal to 3. \textbf{Amazon Dataset\footnote{https://jmcauley.ucsd.edu/data/amazon/}}~\cite{amzon} is a widely-used benchmark dataset~\cite{din,dat,pnn,yu2019adaptive} and our experiment uses a subset Fashion following~\cite{din}.
We take the items with a rating of greater than 3 as positive and the rest as negative. \textbf{Alibaba Dataset\footnote{https://tianchi.aliyun.com/dataset/dataDetail?dataId=56}}~\cite{taobao} is a Taobao ad click dataset.
For the MovieLens and Amazon datasets, following previous work~\cite{inttower}, we divide the train, validation, and test sets by user interaction time in the ratio of 8:1:1. For the Alibaba dataset, we divide the datasets according to the official implementation~\cite{din}, and the data from the previous seven days are used as the training and validation samples with 9:1 ratio, and the data from the eighth day are used for test.

The area under the ROC curve (\textbf{AUC}) measures the probability that the model will assign a higher score to a randomly selected positive item than to a randomly selected negative item.
\textbf{Logloss} is a widely used metric in binary classification to measure the distance between two distributions.

\subsection{Competing Models}
\label{appendix_models}
\textbf{Collaborative Models}: \textbf{Wide\&Deep} combines linear feature interactions (wide) with nonlinear feature learning (deep). \textbf{DeepFM} integrates a Factorization Machine with Wide\&Deep, minimizing feature engineering. \textbf{DCN} enhances Wide\&Deep with a cross-network to capture higher-order interactions. \textbf{AutoInt} uses Multi-head Self-Attention for feature interaction. \textbf{PNN}, \textbf{xDeepFM}, and \textbf{FiBiNET} all serve as strong baselines.

\textbf{Semantic Models}: \textbf{P5} transforms recommendation into text generation with a T5 base, while \textbf{CTR-BERT}, an Amazon model, leverages BERT towers for semantic prediction. \textbf{P-Tab} employs pre-training with Masked Language Modeling (MLM) on recommendation datasets, then fine-tunes for prediction.

% \subsection{Competing Models}
% \label{appendix_models}
% \textbf{Collaborative Models}:
% \textbf{Wide\&Deep}~\cite{widedeep} has been widely used in industry, which contains a wide part and deep part, where the wide part handles the manually designed cross product features while the deep part automatically extracts nonlinear relations among features.  
% \textbf{DeepFM}~\cite{guo2017deepfm} imposes a Factorization Machine as ``wide'' module in Wide\&Deep saving feature engineering jobs. \textbf{DCN}~\cite{dcn} modifies the wide part of the Wide\&Deep model with a cross-network to better learn high-order feature interaction. \textbf{AutoInt}~\cite{autoint} employs Multi-head Self-Attention to automatically build high-order features, which acts as a strong collaborative-based baseline. Additionally, several strong baselines are considered, including 
% \textbf{PNN}~\cite{pnn}, \textbf{xDeepFM}~\cite{xdeepfm}, and \textbf{FiBiNet}~\cite{fibinet}.

% \noindent\textbf{Semantic Models}: \textbf{P5}~\cite{p5} is a semantic-based recommendation model that converts various recommendation tasks into text generation tasks by prompt learning, which uses T5~\cite{t5} as the base model. \textbf{CTR-BERT}~\cite{ctr-BERT} is a semantic two-tower model proposed by Amazon, which adopts two-tower BERT~\cite{BERT} and feeds the semantic information of user and item separately to get the prediction score.
% \textbf{P-Tab}~\cite{ptab} conducts MLM pre-training task on the training set, followed by fine-tuning on downstream score prediction tasks. 

\section{Hyperparameter Analysis}
\subsection{The Impact of Contrastive Learning Temperature Coefficient}
To explore the effect of different temperature parameters in the cross-modal knowledge alignment contrastive learning, we implement experiments on MovieLens and Amazon datasets, and the results are in Figure~\ref{fig:ablation}(a). From the results we can get the following observations: 1) The temperature coefficient in contrastive learning has an obvious impact on the performance. As the temperature coefficient increases, the performance will have a tendency to improve 
first and then decrease, indicating that increasing the coefficient within a certain range is beneficial to improve the performance. 2) For both MovieLens and Amazon datasets, the optimal temperature coefficient is below 1 in our experiments, which has also been verified in previous work~\cite{clip,filip}.

\subsection{The Impact of First Stage  Batch Size}
\begin{figure}[htbp]
\vspace{-2.5em}
	\centering
	\setlength{\belowcaptionskip}{-0.3cm}
	\setlength{\abovecaptionskip}{0cm}
	\subfigure[\small{Temperature Coefficient}]{
		\includegraphics[scale=0.145]{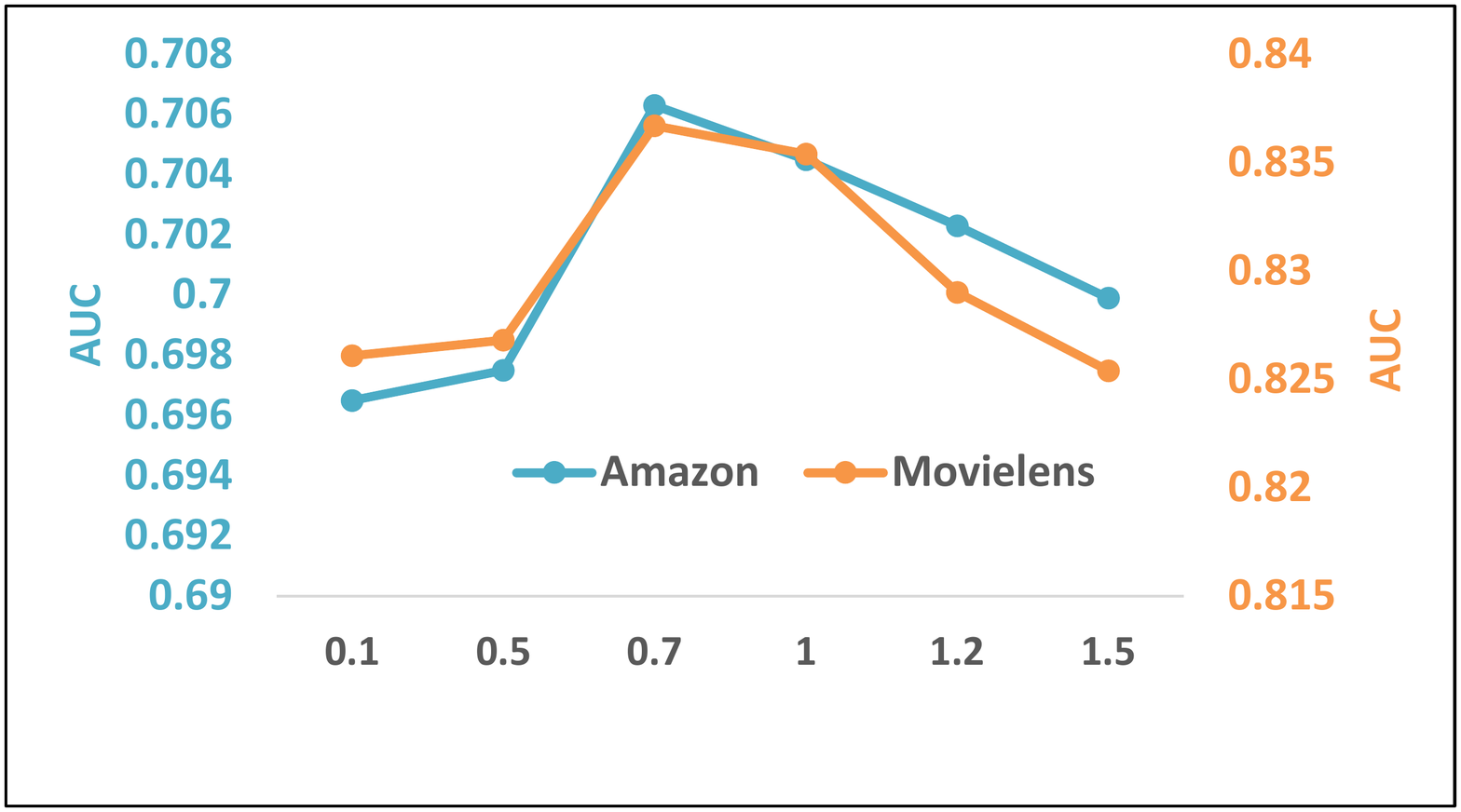}}
	\hspace{-2em}
	\subfigure[\small{Batch sizes}]{
		\includegraphics[scale=0.145]{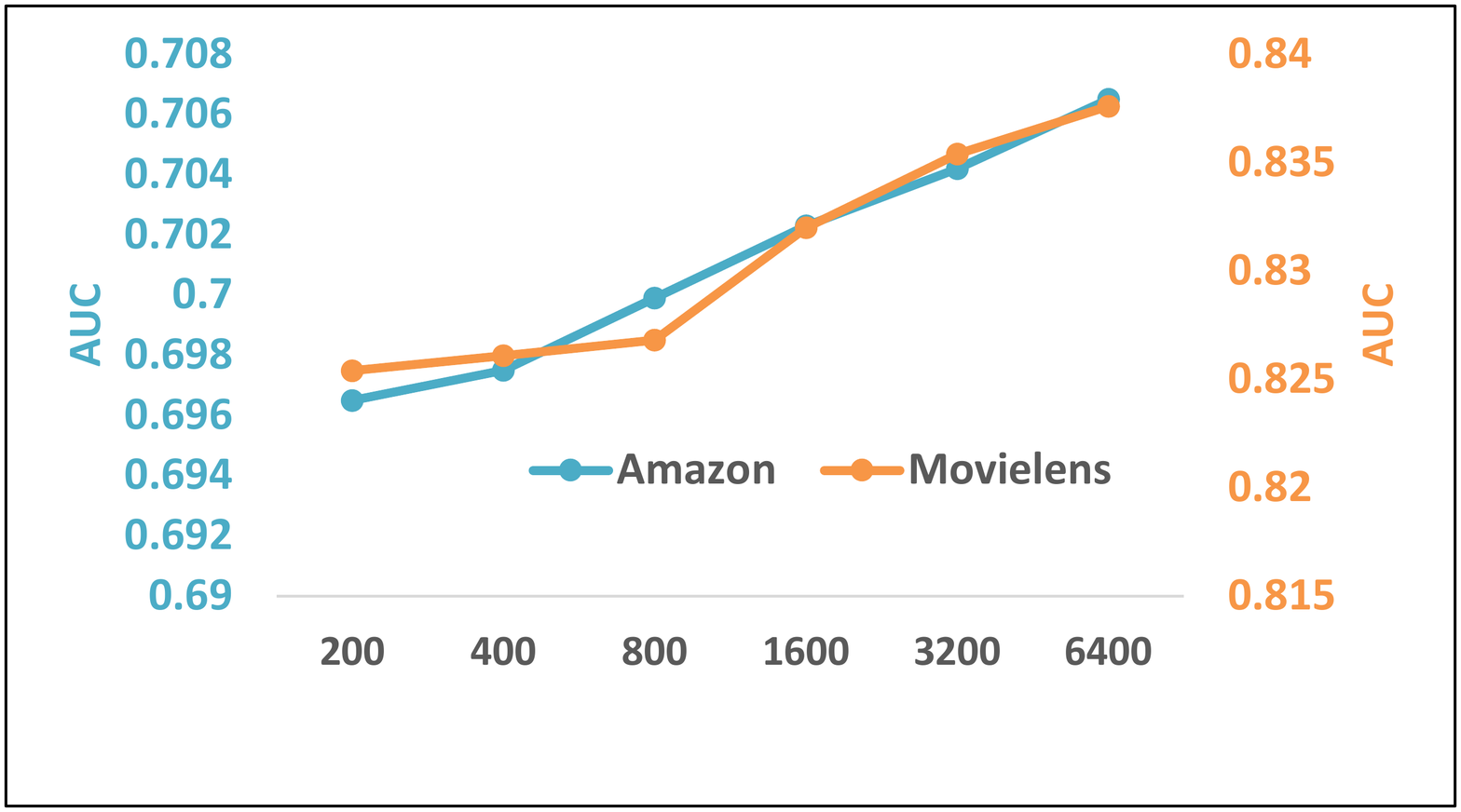}}
% 	\subfigure[\small{Alibaba}]{
% 		\label{fig:ablation:ali} 
% 		\includegraphics[height=80]{figure/ablation3.pdf}}
	\caption{\small{Influence of different contrastive learning temperature coefficient and batch sizes.}}
	\label{fig:ablation}
	\vspace{-1.0em}
\end{figure}

We also explore the impact of different batch sizes, and the results are shown in Figure~\ref{fig:ablation}(b). We can observe that as the batch size increases, the performance is also improved on both datasets, which indicates that increasing the batch size during the contrastive learning pre-training is conducive to achieving better cross-modal knowledge alignment effect and improving the prediction accuracy.

%对比学习不同temper，batch size 
%此处应该画两个图

\end{document}